%% file: main.tex
\documentclass[manuscript]{acmart}

\AtBeginDocument{%
  \providecommand\BibTeX{{%
    \normalfont B\kern-0.5em{\scshape i\kern-0.25em b}\kern-0.8em\TeX}}}

\copyrightyear{2024} 
\acmYear{2024} 
\setcopyright{acmlicensed}\acmConference[CHI '24]{Proceedings of the CHI Conference on Human Factors in Computing Systems}{May 11--16, 2024}{Honolulu, HI, USA}
\acmBooktitle{Proceedings of the CHI Conference on Human Factors in Computing Systems (CHI '24), May 11--16, 2024, Honolulu, HI, USA}
\acmDOI{10.1145/3613904.3642213}
\acmISBN{979-8-4007-0330-0/24/05}

\copyrightyear{2024}
\acmYear{2024}
\setcopyright{acmlicensed}\acmConference[CHI '24]{CHI Conference on Human Factors in Computing Systems}{May 11-16, 2024}{Hawaii, USA}
\acmBooktitle{CHI Conference on Human Factors in Computing Systems (CHI '24), May 11-16, 2024, Hawaii, USA}
\acmPrice{15.00}
\acmISBN{978-1-4503-XXXX-X/18/06}

\usepackage{xcolor}
\usepackage{subcaption}
\usepackage{microtype}
\usepackage{array}
\usepackage{enumitem}
\usepackage{colortbl}
\usepackage{graphicx}
\usepackage{multirow}
\usepackage[]{algorithm2e}
\usepackage{multirow}
\usepackage{graphicx}
\usepackage{calc}
\usepackage[
  separate-uncertainty = true,
  multi-part-units = repeat
]{siunitx}
\usepackage{url}

\def\markup{0}

\if\markup1
\usepackage[normalem]{ulem}
  \definecolor{myblue}{rgb}{0,0,0.75}
  \newcommand{\rv}[1]{{\leavevmode\color{myblue}#1}}
  \newcommand{\st}[1]{{\sout{#1}}}
\else
  \newcommand{\rv}[1]{#1}
\newcommand{\st}[1]{}
\newcommand{\sout}[1]{}
\fi

\newcommand{\forceindent}{\leavevmode{\parindent=1em\indent}}

\newcolumntype{\$}{>{\global\let\currentrowstyle\relax}}
\newcolumntype{^}{>{\currentrowstyle}}

\acmSubmissionID{9641}

\begin{document}


\title{FetchAid: Making Parcel Lockers More Accessible to Blind and Low Vision People With Deep-learning Enhanced Touchscreen Guidance, Error-Recovery Mechanism, and AR-based Search Support}

\author{Zhitong Guan}
\authornote{Both authors contributed equally to this research.}
\affiliation{
  \institution{Computational Media and Arts Thrust}
  \institution{The Hong Kong University of Science and Technology (Guangzhou)}
  \city{Guangzhou}
  \country{China}
}
\email{klaraztguan@ust.hk}

\author{Zeyu Xiong}
\authornotemark[1]
\affiliation{
  \institution{Computational Media and Arts Thrust}
  \institution{The Hong Kong University of Science and Technology (Guangzhou)}
  \city{Guangzhou}
  \country{China}
}
\email{zxiong666@connect.hkust-gz.edu.cn} 

\author{Mingming Fan}
\orcid{0000-0002-0356-4712}
\affiliation{%
  \institution{Computational Media and Arts Thrust}
  \institution{The Hong Kong University of Science and Technology (Guangzhou)}
  \city{Guangzhou}
  \country{China}
}
\affiliation{%
  \institution{Division of Integrative Systems and Design \& Department of Computer Science and Engineering}
  \institution{The Hong Kong University of Science and Technology}
  \country{Hong Kong SAR, China}
}
\authornote{Corresponding Author}
\email{mingmingfan@ust.hk}

\renewcommand{\shortauthors}{Guan, Xiong and Fan}

\begin{abstract}

 Parcel lockers have become an increasingly prevalent last-mile delivery method. Yet, a recent study revealed its accessibility challenges to blind and low-vision people (BLV). Informed by the study, we designed FetchAid, a standalone intelligent mobile app assisting BLV in using a parcel locker in real-time by integrating computer vision and augmented reality (AR) technologies. FetchAid first uses a deep network to detect the user's fingertip and relevant buttons on the touch screen of the parcel locker to guide the user to reveal and scan the QR code to open the target compartment door and then guide the user to reach the door safely with AR-based context-aware audio feedback. Moreover, FetchAid provides an error-recovery mechanism and real-time feedback to keep the user on track. We show that FetchAid substantially improved task accomplishment and efficiency, and reduced frustration and overall effort in a study with 12 BLV participants, regardless of their vision conditions and previous experience.
  
  
  
  

\end{abstract}



\maketitle

\input{Doc/1-introduction}

\input{Doc/2-related-work}

\input{Doc/3-design}

\input{Doc/4-implementation}
\input{Doc/5-technical}
\input{Doc/6-userstudy}
\input{Doc/7-results}

\input{Doc/8-discussion}
\input{Doc/9-conclusion}



\begin{acks}
This work is partially supported by 1) 2024 Guangzhou Science and Technology Program City-University Joint Funding Project (PI: Mingming Fan); 2) 2023 Guangzhou Science and Technology Program City-University Joint Funding Project (Project No. 2023A03J0001) ; 3) Guangdong Provincial Key Lab of Integrated Communication, Sensing and Computation for Ubiquitous Internet of Things (No.2023B1212010007).
\end{acks}

\bibliographystyle{ACM-Reference-Format}
\bibliography{main.bib}

\newpage
\appendix
\input{Doc/appendix}


\end{document}

%% file: Doc/1-introduction.tex
\section{Introduction}

\begin{figure}[tbh!]
    \centering
    \includegraphics[width=0.8\linewidth]{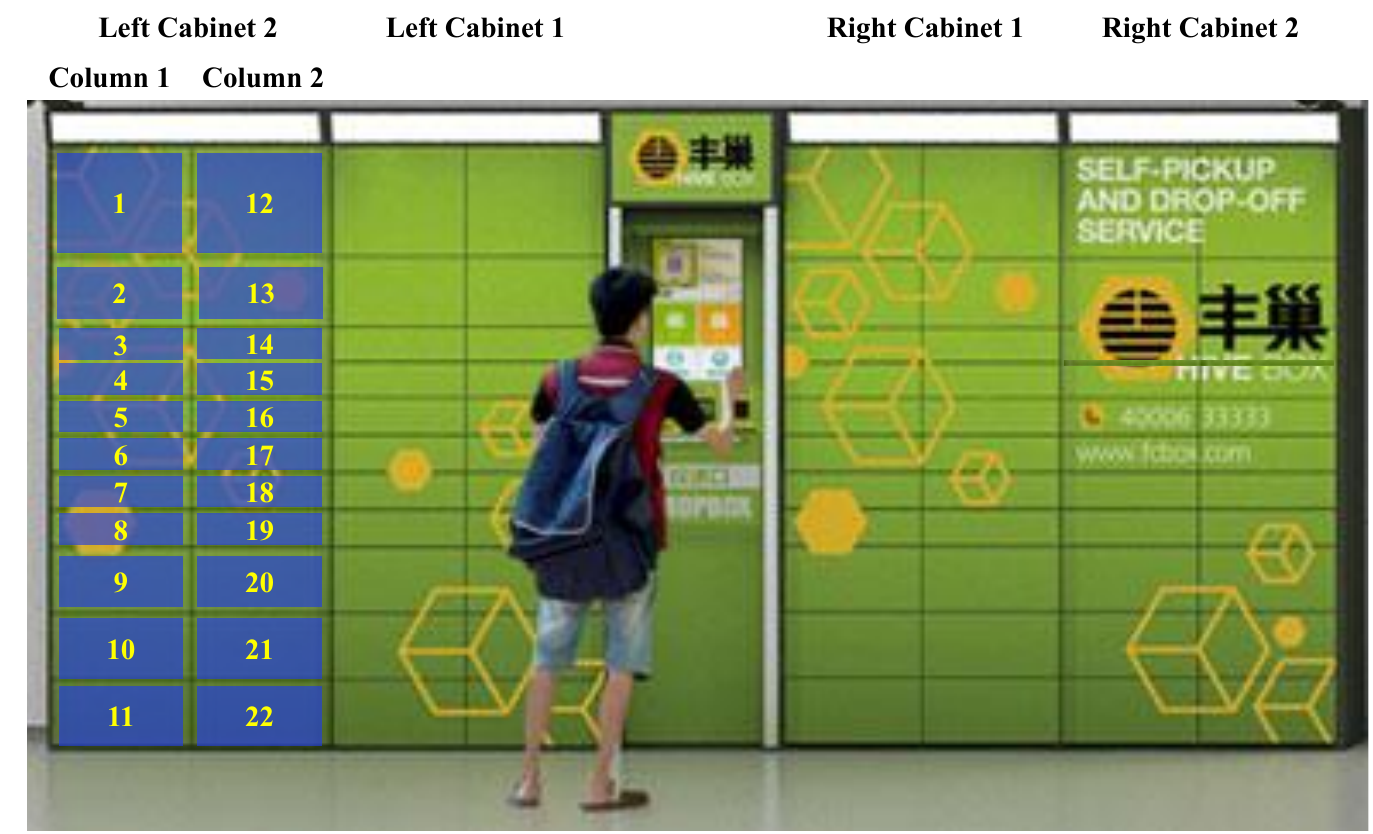}
    \caption{A Standard HiveBox parcel locker in China. Each parcel locker is modular and consists of several cabinets to the left and right of the center touchscreen interface. Each cabinet consists of two columns and each column contains 11 compartments. The dimensions of each compartment are standardized.}
    \label{fig:hivebox}
    \Description{Figure 1 shows a standard HiveBox parcel in China. Each parcel locker is modular and consists of several cabinets to the left and right of the center touchscreen interface. Each cabinet consists of two columns and each column contains 11 compartments. The dimensions of each compartment are standardized.}
\end{figure}

The growth of online shopping has led to a significant increase in daily package deliveries, posing multiple challenges for the traditional express delivery industry, such as sustainability, costs, an aging workforce, and time pressure \cite{boysen_fedtke_schwerdfeger_2020}. Limited workforce availability and transportation costs have added further strain on logistic systems \cite{capacity_schöder_ding_campos_2016}. Courier services often encounter failed first-time home deliveries, which can occur in 12 to 60\% of cases worldwide, resulting in wasted time and resources \cite{lastmileSong}. Consequently, efficient alternative delivery methods, such as unattended deliveries and customer self-service options, have gained traction \cite{boysen_fedtke_schwerdfeger_2020}, prompting the application of various last-mile delivery technologies \cite{Winkenbach, lastmileDELLAMICO20121505, lastmilePoeting, boysen_fedtke_schwerdfeger_2020}. Among them, parcel lockers (see Figure \ref{fig:hivebox} for an example) have become a widely adopted and effective solution across numerous countries \cite{lastmileYael,capacity_schöder_ding_campos_2016}, such as the US's Amazon Hub Locker, Germany's DHL Packstation, the UK's InPost, and China's HiveBox \cite{ Fengchao, rohmer2020guide}~\footnote{\url{https://www.dhl.de/en/privatkunden/pakete-empfangen/an-einem-abholort-empfangen/packstation/empfangen-packstation.html}}.
Typically, parcel lockers consist of a touchscreen for entering retrieval information and compartments for secure package storage. 


Retrieving packages from parcel lockers is highly visually demanding and poses challenges for Blind and Low Vision (BLV) people \cite{chi22blvPackage}. A recent study revealed that BLV struggles with navigating complex onscreen workflow, recovering from incorrect onscreen actions, and avoiding pop-up advertisements \cite{chi22blvPackage}. Even if they successfully navigate the touchscreen, they must still locate the open compartment and safely approach it without tripping or colliding with its door \cite{chi22blvPackage}.

Creating an accessible package retrieval experience from parcel lockers for BLV necessitates a novel, integrated solution that concurrently addresses three core challenges: touchscreen interaction flow, object (open compartment) localization, and navigation. While prior research has tackled similar challenges individually in different contexts, such as computer vision and crowdsourcing for camera aiming and touchscreen information \cite{Anhong2016vizlens}, wearable devices for object recognition and localization \cite{Boldu2018FingerReader2, chen2022lisee}, and systems like NavCog for turn-by-turn navigation assistance \cite{jafri2014computer, meshram2019astute,ahmetovic2016navcog,sato2017navcog3}, there is a need for an innovative, systematic solution that holistically addresses all three challenges, without requiring extra hardware and online help.


We introduce FetchAid, a mobile app system that offers real-time guidance and feedback, enabling BLV to retrieve packages from parcel lockers by addressing the three aforementioned challenges. 
FetchAid assists BLV in two phases: Touchscreen Interaction and Open Door Searching. 
During the Touchscreen Interaction phase, parcel locker user interfaces present unique challenges, such as distractions from advertisements and extraneous buttons, which may lead to errors \cite{chi22blvPackage}. Without a screen reader or Braille, BLV struggles to verify its actions \cite{chi22blvPackage}. 


Previous work on touchscreen accessibility for BLV, though innovative, has limitations. Hardware solutions like Talking Fingertip \cite{Vanderheiden1996UseOA}, Talking Tactile Tablet \cite{ttt}, and Touchplates \cite{touchplates} improve accessibility but entail high costs and practical issues. Software solutions like NavTap \cite{navtap}, Slide Rule \cite{sliderule}, and Access Overlays \cite{accessoverlays} enhance usability but demand access to the device's internal system, which can be problematic. Additionally, applications such as VizLens \cite{guo2016vizlens} and StateLens \cite{guo2019statelens} employ computer vision and crowd-sourcing for interface accessibility. However, these lack comprehensive evaluations under varying light conditions and instant error detection or recovery mechanisms.

FetchAid steps in to address these challenges. It tracks user finger movements relative to the target button, offers real-time vocal guidance and feedback, and incorporates an error-recovery mechanism that, upon detecting an error, guides the user back to the previous page. Once the correct button is activated, FetchAid scans the QR code, unlocking the compartment door and eliminating the need for further user input.


After unlocking the door, users face additional challenges: locating the door despite small openings or occlusions and safely navigating to it without tripping or colliding with the open door \cite{chi22blvPackage}. Previous research has investigated helping BLV people reach surrounding targets \cite{chen2022lisee}, and recognizing objects \cite{lee2019hands, jafri2014computer, manduchi2014last}. However, these approaches rely on wearable devices, remote workers, or vision-based algorithms, which have limitations in availability and robustness. Another challenge is guiding users to the identified door while being aware of its 3D location to avoid accidental collisions. Additionally, the navigation tool must provide high-frequency feedback, making remote-worker assistance impractical. In the Open Door Search phase, FetchAid accurately determines the open door's location using touchscreen text readings, a more reliable method than vision-based alternatives. Utilizing Augmented Reality (AR) to localize users relative to the target door in a closed-loop fashion, FetchAid then provides real-time, intuitive voice guidance to help BLV people reach the door.

We conducted a technical evaluation to validate the functionality of key technical components of FetchAid, and a user study with 12 BLV participants to assess the capability of FetchAid in making parcel locker package fetching accessible to BLV users. Quantitative and qualitative results indicate that FetchAid has increased success rates, efficiency, and reduced task load for BLV users.

This paper makes the following contributions:

\begin{itemize}
\item A standalone mobile assistive technology, FetchAid, to help BLV people overcome the challenges of the interaction flow on a parcel locker touchscreen and recover from possible mistakes and assist them to navigate safely to the open compartment door with AR-based context-aware voice guidance.

\item Technical and user evaluations that show the performance and user experience of FetchAid in assisting BLV people with fetching packages.
\end{itemize}

%% file: Doc/2-related-work.tex
\section{BACKGROUND AND RELATED WORK}
We first present background information about parcel lockers and then report prior work related to three key challenges of fetching packages from parcel lockers for BLV people: \textit{making touchscreen more accessible to BLV people}; \textit{making objects finding more accessible to BLV people}; and \textit{making navigation more accessible to BLV people}. 
\subsection{Background of Package Delivery and Parcel Lockers}
\begin{figure*}[bth!]
    \centering
    \includegraphics[width=\textwidth]{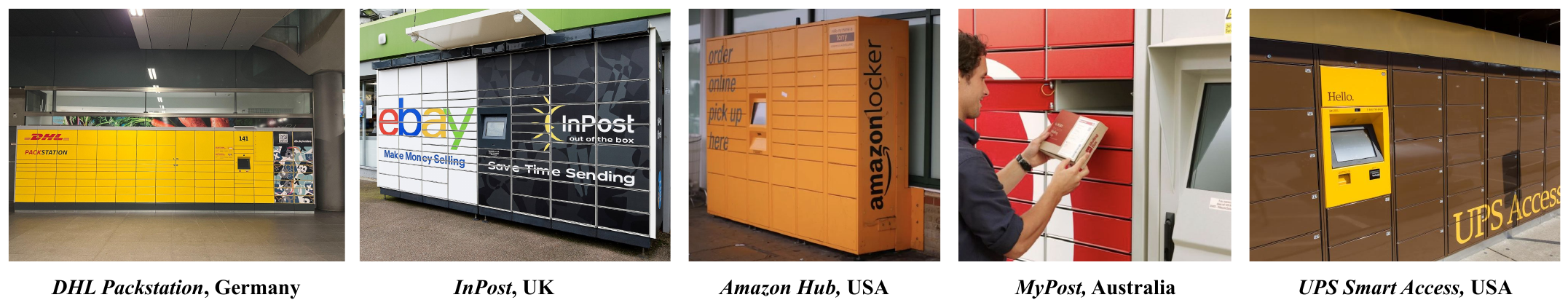}
    \caption{Parcel lockers worldwide.}
    \label{fig:worldwideparcel}
    \Description{Figure 2 shows 5 examples of parcel lockers worldwide, which are: Parcel Lock in Germany, InPost in UK, Amazon Hub in USA, MyPost in Australia, and UPS Smart Access in USA.}
\end{figure*}

Global parcel volume has reached 159 billion, with projections to hit 216 to 300 billion by 2027~\footnote{\rv{\url{https://www.pitneybowes.com/content/dam/pitneybowes/us/en/shipping-index/22-pbcs-04529-2021-global-parcel-shipping-index-ebook-web-002.pdf}}}. "Last-Mile Delivery," the final step in parcel delivery from a hub to its destination, has been a bottleneck in express delivery efficiency~\footnote{\url{http://hb.ifeng.com/a/20200509/14209060_0.shtml}}. 
While numerous innovative last-mile delivery concepts have been proposed~\cite{boysen2021last}, parcel lockers have become one of the most widely adopted. Fig. \ref{fig:worldwideparcel} showcases examples of parcel lockers in the US, Germany, the UK, and Australia.

Parcel lockers enable couriers and users to store and retrieve packages, typically featuring metal cabinets with various compartments. To pick up packages, users interact with a touchscreen interface to open the target compartment and collect the parcel, a process consistent across regions~\footnote{\url{https://auspost.com.au/mypost/how-to/deliveries.html}}~\footnote{\url{https://www.dhl.de/en/privatkunden/pakete-empfangen/an-einem-abholort-empfangen/packstation/empfangen-packstation.html}}. 
As all researchers in this study reside in China, we selected HiveBox as the example parcel locker.

\begin{figure*}[tbh!]
    \centering
    \includegraphics[width=\linewidth]{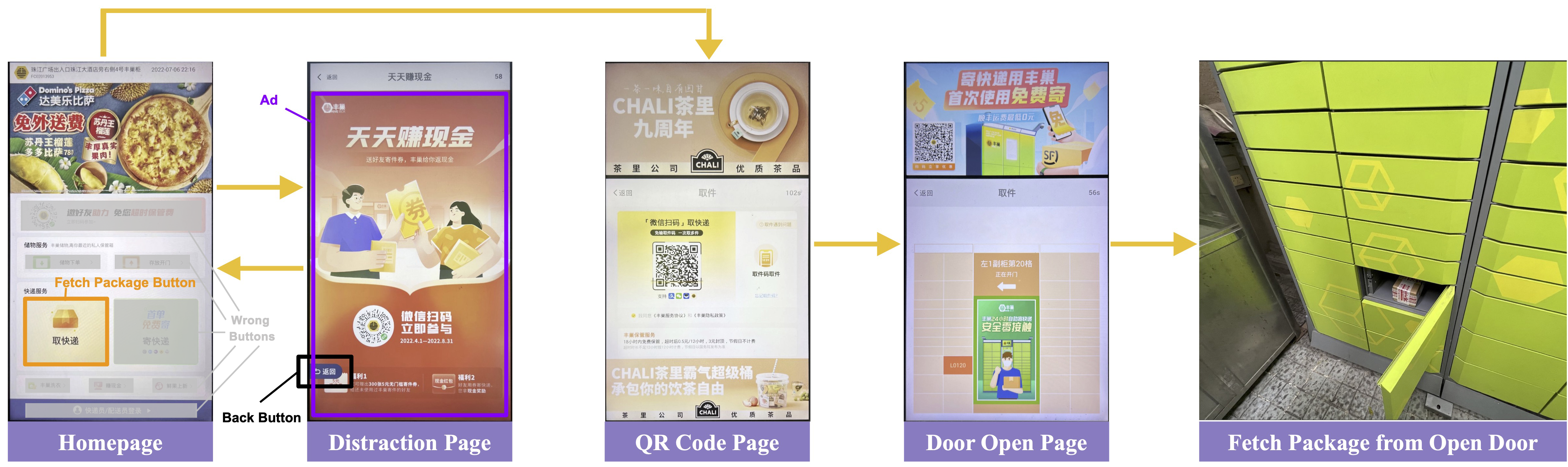}
    \caption{Task flow of the HiveBox parcel locker touchscreen. The user starts with the Homepage. To fetch a package from a HiveBox, the user needs to first tap the ``Fetch Package'' button on the homepage, which consequently reveals the QR Code Page. The user needs to scan the QR code to unlock the door. The QR code will not be revealed if the user clicks the wrong button, which instead reveals a Distraction Page (e.g. an advertisement and a back button), and the user needs to press the ``back button'' to return to the previous page. The compartment location is shown on the Door Open Page after the QR code is scanned for user authentication.}
    \label{fig:pipeline}
    \Description{Figure 3 shows the task flow of the HiveBox parcel locker touchscreen. The user starts with the Homepage. To fetch a package from a HiveBox, the user needs to first tap the ``Fetch Package'' button on the homepage, which consequently reveals the QR Code Page. The user needs to scan the QR code to unlock the door. The QR code will not be revealed if the user clicks the wrong button, which instead reveals a Distraction Page (e.g. an advertisement and a back button). The compartment location is shown on the Door Open Page after the QR code is scanned for user authentication.}
     \label{fig:hiveboxinteractionflow}
\end{figure*}

HiveBox, China's leading parcel locker with over 400,000 sets deployed in 2020 and a 70\% market share~\footnote{\url{https://finance.sina.com.cn/roll/2020-05-06/doc-iircuyvi1621900.shtml}}, consists of multiple cabinets, each containing 22 compartments and a touchscreen interface for package deposit and retrieval (see Fig.~\ref{fig:hivebox}) \cite{Fengchao}. After a courier deposits a package, HiveBox sends the recipient a pickup notification containing the locker location. Fig. \ref{fig:hiveboxinteractionflow} shows the user interaction flow at the locker. The user must select "Fetch Package" on the touchscreen to access the package, choosing either QR code scanning in WeChat or an eight-digit pickup code for verification. Once verified, a compartment door opens for package collection. 

Despite their prevalence, most parcel lockers lack accessibility features like voice feedback, screen readers, or Braille keypads, underscoring the need to improve accessibility \cite{chi22blvPackage}. Although some parcel lockers feature Braille keypads and headphone jacks \cite{Luís_Martins_Caldeira_Soares_2022, TalkingLocker}, they have limitations beyond cost. Firstly, less than 10\% of the blind and low-vision population read Braille \cite{Braille_Literacy_Crisis}. Secondly, headphone jacks present mobility and safety issues due to limited wire length, especially as the dimensions of parcel lockers typically exceed wire length.

\subsection{Making Touchscreen Interaction More Accessible to BLV people} 


One line of prior work made touchscreens more accessible to BLV people by adding custom devices or modifying hardware. For instance, Talking Fingertip allowed users to interact with a kiosk's touchscreen using an external momentary switch \cite{Vanderheiden1996UseOA}. Talking Tactile Tablet combined a touch-sensitive pad with a graphical overlay for audible and tactile feedback \cite{ttt}. Touchplates used custom hardware overlays and software to provide tactile feedback \cite{touchplates}. TouchA11y provided an intelligent robot to replace human fingers \cite{10.1145/3544548.3581254}. However, these approaches can be costly or impractical for end-users due to the need for hardware modifications or additional equipment.



Another line of prior work aimed to make touchscreen devices more accessible by interfacing with their software. For instance, NavTap \cite{navtap} and Slide Rule \cite{sliderule} assist BLV people in interacting with multi-touch screen applications using audio-based input gestures. Access Overlays focused on helping BLV people identify elements on large touchscreens \cite{accessoverlays}. However, these approaches require access to the device's software, which is often unavailable to end-users.


Another prevalent line of prior work made touchscreen and non-touchscreen devices more accessible using smartphones. For instance, VizLens \cite{Anhong2016vizlens} and StateLens \cite{guo2019statelens} are mobile solutions that employ computer vision and crowd-sourcing to make appliance interfaces accessible to BLV people. Similar specialized computer vision systems assist BLV users in reading LCD appliance panels \cite{fusco2014using, tekin2011real, morris2006clearspeech}. These approaches primarily focus on recognizing interface elements. However, BLV people often face challenges when accidentally touching unwanted elements on parcel locker touchscreens (e.g., advertisement button) without realizing it \cite{chi22blvPackage}. As a result, they struggle to recover from these errors \cite{chi22blvPackage}. Previous work has not proposed robust error detection and recovery mechanisms for BLV users. In this study, we address this challenge by detecting errors and helping BLV people recover from them.


\subsection{Making Objects Finding More Accessible to BLV people}\label{sec: Object Framing in Help with Camera Aiming}
When BLV people successfully navigate the touchscreen interaction challenges, the target compartment door containing the package will open, potentially making a clicking sound. However, parcel lockers are often located in noisy, high-traffic areas like shopping malls or subways, making the sound difficult to perceive. Even if the sound is audible, it may only offer a general direction of the open compartment rather than its precise location within the parcel locker \cite{chi22blvPackage}. 


Previous work explored helping BLV people locate surrounding objects \cite{chen2022lisee}, improve object recognition \cite{lee2019hands, jafri2014computer}, and navigate the last meter to the target object \cite{manduchi2014last}. Many of these approaches depend on computer vision algorithms and are susceptible to object-out-of-frame situations. To address framing and camera aiming challenges, techniques such as VizWiz::LocateIt used remote workers' annotations \cite{vizwiz}, EasySnap provided an accessible photo album for non-visual review and sharing \cite{jayant}, and \cite{vazquez} developed a system for camera aiming using RoI proposals and audio feedback. However, these approaches often rely on remote human workers or computer-vision techniques that need numerous training images under various conditions. In contrast, Our assistive tool sidesteps this problem by analytically computing the location of the open door relative to the user \textit{in the world frame} by understanding the text prompt shown on the touchscreen and the structure of the locker. 


\subsection{Making Navigation More Accessible to BLV people}
\label{subsec:navigation}
Once the open compartment door is located, BLV people must be guided to it. Prior work explored computer vision-based approaches for indoor mapping \cite{li2018vision, meshram2019astute} or used Bluetooth or ultra-wideband wireless positioning technology for precise navigation by providing nearby points-of-interest (POI), including accessibility issues (e.g., stairs ahead), or turn-by-turn information \cite{precisionNAV, ahmetovic2016navcog,sato2017navcog3}. Recently, AR-based smartphone apps demonstrated the ability to recognize and track real-world objects as visual cues for BLV people's outdoor navigation \cite{coughlan2017ar4vi, kaul2021mobile}. \citet{ferrand2018augmented} proposed an AR-based system using 3D audio for outdoor navigation, and LineChaser is an AR app for guiding BLV people in queues \cite{LineChaser}. Informed by this line of work, we leverage the Visual-Inertial Odometry (VIO) capability of mobile AR \cite{forster2016manifold, zhang2020dex} to localize users in real-time, accurately guiding them to the open compartment door. Additionally, we designed context-aware voice guidance based on AR readings to help BLV people avoid accidents with the open door, addressing a key safety concern identified in a recent study \cite{chi22blvPackage}.

%% file: Doc/3-design.tex
\section{Design of FetchAid}

Drawing from prior work, particularly a recent study revealing key challenges BLV users face when using parcel lockers \cite{chi22blvPackage}, we designed an assistive smartphone app, FetchAid, to make package retrieval from parcel lockers more feasible for BLV people. Based on the literature, we identified seven design considerations (\textbf{D}):

\textbf{D1: Guide BLV people to the target button on the parcel locker's touchscreen.} Since most BLV people lack experience with parcel lockers and a mental model of the touchscreen interface, it is crucial to guide them toward the target button. Results from a study by \citet{guo2016vizlens} indicate that guiding users to the target button achieves a higher success rate than providing feedback on which visual element their finger is hovering over. Specifically, the guidance should instruct users to move in certain directions and tap the touchscreen when they reach the target button.

\textbf{D2: Offer real-time feedback on interface operation status.} BLV parcel locker users require constant feedback on their operation status, including both failures and successes. They may be unsure if they've tapped the correct button and entered the package retrieval page, and they may accidentally tap other elements, leading to irrelevant pages such as advertisements \cite{chi22blvPackage}. In these cases, BLV people struggle to recognize they are on the wrong page and have difficulty returning to the homepage due to limited feedback and complex interfaces. FetchAid addresses these needs by intelligently recognizing page changes and providing voice feedback to inform users of their current page.

\textbf{D3: Implement an error-recovery guide for irrelevant page entries in FetchAid.} Building on D2, once users realize they've entered an irrelevant page, they need guidance to return to the previous page or homepage. By guiding users back to the homepage, they can restart the process with minimal effort and frustration.

\textbf{D4: Enable streamlined QR code scanning for automatic compartment door opening, rather than requiring users to input a pickup code.} When users reach the "Fetch Package" page, they can either manually input the pickup code or scan the parcel locker's touchscreen QR code to unlock the compartment containing the package. While scanning a QR code is simple, it necessitates navigation to a dedicated QR-scanning app like WeChat. On the other hand, the pickup code method bypasses the need for smartphone action but forces BLV users to screen-read an eight-digit code. Furthermore, interacting with a touchscreen keyboard poses another challenge for BLV people due to the lack of a physical keypad on a parcel locker. FetchAid streamlines the door-unlocking process by guiding users to point at the QR code and then automatically triggering the scanning process within the same app, once the door-opening QR code becomes visible.

\textbf{D5: Guide BLV users to the open compartment door.}  After successfully unlocking the compartment door by tapping the "Fetch Package" button, locating the open door remains error-prone for BLV people. The "click" sound produced when the door unlocks may provide a general direction but \st{requires extra exploration and} \rv{they need guidance to get to the exact opened door as there were many compartments in any given direction~\cite{chi22blvPackage}.} \rv{And audio cue} is unreliable in noisy outdoor environments. FetchAid aims to acquire accurate user and compartment location coordinates, offering real-time guidance by tracking these coordinates using AR-based VIO techniques.

\textbf{D6: Assist BLV people users in avoiding collisions while navigating to the open door.} When the target door is at the top, bottom, far right, or far left of the parcel locker, BLV people may accidentally bump into, trip over, or inadvertently close the open door. FetchAid notifies users if the door is at their feet or head level and alerts them to step away if they get too close to the cabinet, preventing collisions.

\textbf{D7: Detect out-of-frame situations and guide users to properly position the camera.} Prior mobile assistive technology emphasizes the importance of helping users aim the camera at the interface center. Without this feature, users may not realize the interface is out of view and continue using the system. FetchAid should assist users in better aiming the camera. It detects whether the interface's center is within the phone's camera view, providing feedback like "Please move the phone to the right" to help users adjust the camera angle.


%% file: Doc/4-implementation.tex
\begin{figure*}[tbh!]
  \centering
  \includegraphics[width=\linewidth]{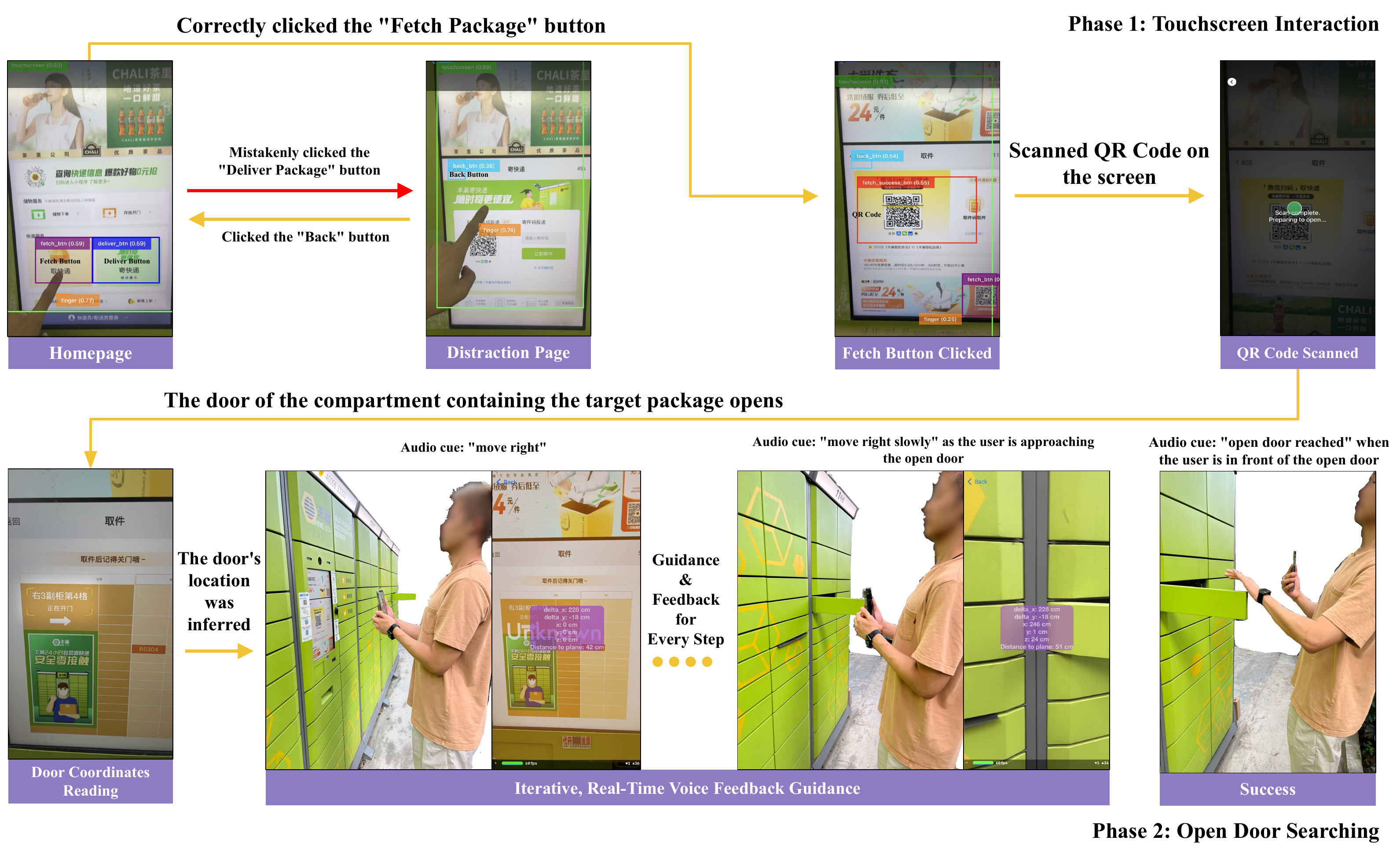}
  \caption{FetchAid system overview: The system assists BLV in two phases, \rv{Touchscreen Interaction and Open Door Searching.}\st{with images featuring green backgrounds taken from the parcel locker's touchscreen, and those with yellow backgrounds from FetchAid's interface.} In the Touchscreen Interaction Phase, FetchAid detects the overview page, tracks the user's fingertip (bounded by the orange-label box), and guides the user to tap "Fetch Package" (bounded by the violet-label box). If the user accidentally taps "Deliver Package" (blue box), FetchAid directs them to the "Back" button (cyan box). After tapping "Fetch Package," a QR code (red box) appears, is scanned, and a target door opens, initiating the Open Door Searching Phase. FetchAid then provides real-time voice feedback based on the user's estimated position from ARKit, guiding them to the open door.}
  \label{fig:overview}
  \Description{Figure 4 illustrates the system overview of FetchAid, displaying FetchAid interfaces. The system assists BLV in two phases: Touchscreen Interaction and Open Door Searching. In the Touchscreen Interaction Phase, FetchAid detects the overview page, tracks the user's fingertip (bounded by the orange-label box), and guides the user to tap "Fetch Package" (bounded by the violet-label box). If the user accidentally taps "Deliver Package" (blue box), FetchAid directs them to the "Back" button (cyan box). After tapping "Fetch Package," a QR code (red box) appears, is scanned, and a target door opens, initiating the Open Door Searching Phase. FetchAid then provides real-time voice feedback based on the user's estimated position from ARKit, guiding them to the open door.}
\end{figure*}

\section{Implementation of FetchAid}
FetchAid is a standalone smartphone app that doesn't require modifications to existing parcel locker interfaces. While prior work has addressed BLV navigation to target locations (see Section \ref{subsec:navigation}), our focus is on assisting BLV with parcel locker touchscreen interactions. 
FetchAid is designed to facilitate these interactions through a two-step pipeline, as shown in Fig. \ref{fig:overview}. We provide detailed information about FetchAid's system components, with code to be made available upon acceptance for better reproducibility.

\subsection{Touchscreen Interaction Phase}
\begin{figure}[tbh!]
    \centering
    \includegraphics[width=\linewidth]{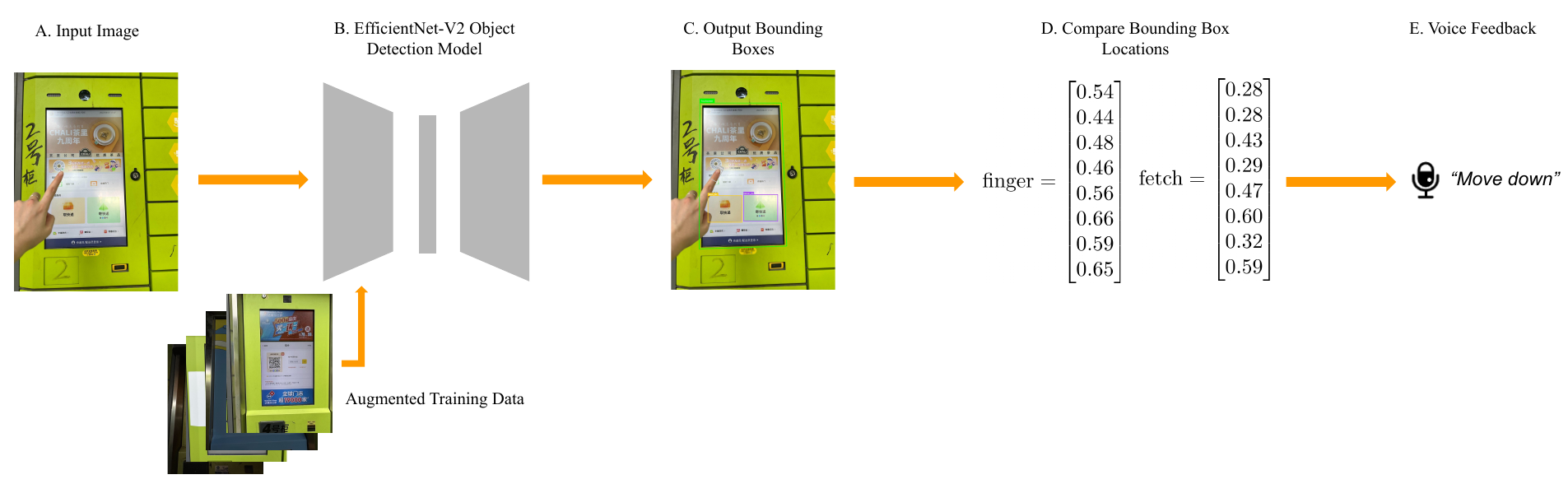}
    \caption{The object detection model in FetchAid for identifying graphical elements in the UI. An EfficientNet-V2-based network outputs the bounding boxes of important graphical elements in the UI and the user's pointing finger, which will be used to output voice feedback to instruct the user to click the desired button.}
    \label{fig:objdetection}
    \Description{Figure 5 describes the object detection model in FetchAid for identifying graphical elements in the UI. An EfficientNet-V2-based network outputs the bounding boxes of important graphical elements in the UI and the user's pointing finger, which will be used to output voice feedback to instruct the user to click the desired button.}
\end{figure}

FetchAid's first phase involves identifying essential graphic elements on the touchscreen, such as buttons, banners, and the QR code needed for unlocking the compartment. The system runs a real-time object detection model on the user's smartphone to track the user's index fingertip and the "Fetch Package" button to guide BLV users accurately (\textbf{D1}). 
The tracking result will be used to generate voice feedback to help users tap the desired button. Throughout this phase, FetchAid continuously verifies if the touchscreen is within the camera's field of view, prompting users to adjust their aim if needed (\textbf{D7}). An illustration of model training and outputting detection is illustrated in Fig. ~\ref{fig:objdetection}.

\subsubsection{Object Detection Model Training} 
We trained a lightweight deep-learning-based object detection model to detect the user's finger and the interface's visual elements in real time. While previous work detected fingers and UI elements by having human experts annotate inaccessible keys on the UI, those studies often assumed fixed camera poses and static UI, which limited the generalizability and versatility of the assistive tool \cite{guo2016vizlens, fusco2014using, tekin2011real, morris2006clearspeech}. 


To collect training data, we manually gathered 473 pictures of touchscreens from 9 different parcel lockers. The images were taken with an iPhone camera held in one hand (either left or right) under varying camera perspectives and lighting conditions. Some pictures included a pointing index finger from the other hand, while others did not. We ensured that all different pages and graphic elements were captured.

The 473 images were labeled using OpenCV's CVAT~\cite{boris_sekachev_2020_4009388} toolkit, with bounding boxes manually drawn for the index fingertip, "Fetch Package" button, "Deposit Package" button, "Back" button, QR code (appearing after tapping the "Fetch" button) page's banner, and the full touchscreen. To improve the detection model's robustness under different lighting conditions and camera perspectives, we augmented the dataset using HSV (Hue, Saturation, Value) colorspace randomization and random affine transformations (rotation, translation, shearing, and scaling). Each image was augmented 10 times, each time with a random combination of the described transformations, resulting in a final dataset of 4730 images. Fig. ~\ref{fig:cvat} in Appendix \ref{appendix} provides examples of the data labeling process and data augmentation.

We divided our dataset into an 8-1-1 train-validation-test split and trained the network for 15 epochs. The training loss stabilized and converged at the end. The model was implemented using TensorFlow's EfficientNet-Lite 2 \cite{tan2019efficientnet} backbone, customized for efficient performance on smartphones. We exported the trained model to TF-Lite format and integrated it into the TensorFlow-Lite framework in iOS.

At inference time, the model runs on the iPhone's native CPU. The resulting object detection model can track the user's finger and the touchscreen's graphic elements in real-time with high accuracy. The detector operates at 10 fps, and in each frame, the model outputs detected elements of interest along with their bounding boxes.

\begin{figure}[tbh!]
    \centering
    \includegraphics[width=\linewidth]{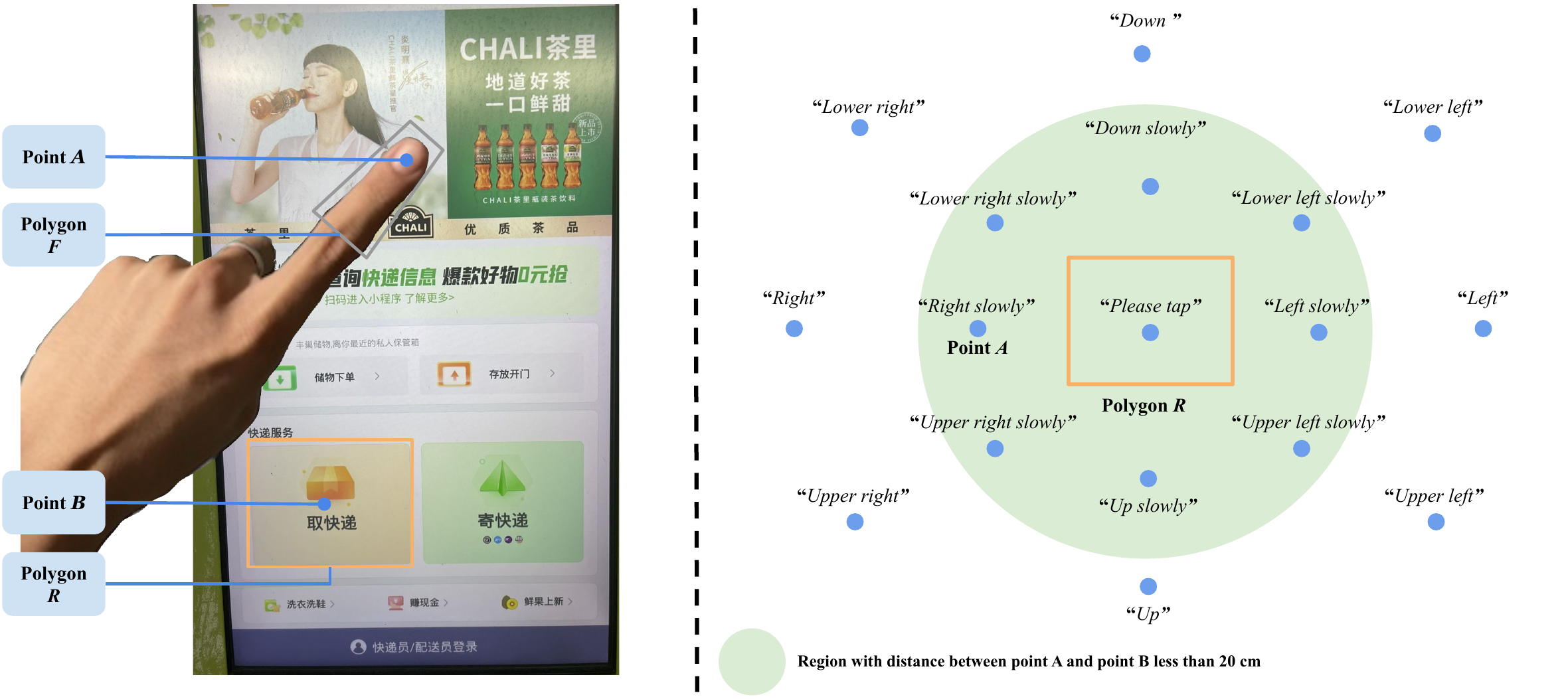}
    \caption{Illustration of Touchscreen Interaction with FetchAid. \textbf{Left:} Point $A$ and Polygon $F$ track the user's fingertip, while Point $B$ and Polygon $R$ track the target button (e.g., "Fetch Package" button). \textbf{Right:} Voice feedback and guidance are provided for different directions based on the relative position between the fingertip (Point $A$) and the target button (Point $B$). When the distance between Point $A$ and Point $B$ is less than 20 cm, users are reminded to move slowly. Distances in this step are heuristically determined using perspective projection.}
    \label{fig:touchingpoint}
    \Description{Figure 6 illustrates the Touchscreen Interaction with FetchAid. The figure on the left shows Point A and Polygon F, which are used to track the user's fingertip. Point B and Polygon R are used to track the target button. The target button in this figure is the ``Fetch Package'' button. The figure on the right illustrates the voice feedback for different directions. FetchAid tells the user to move in these directions according to the relative position between the fingertip (Point A) and the target button (Point B). Note that once the distance between Point A and Point B is less than 20 centimeters, the user will be reminded to move slowly. Distances in this step are heuristically determined using perspective projection.}
\end{figure}
\subsubsection{Navigation across the Touchscreen and Feedback}
Utilizing detected bounding box locations, we developed a policy to efficiently guide users to the "Fetch Package" button. FetchAid provides real-time voice feedback, directing BLV users to position their finger over the target button while holding their phone's camera towards the parcel locker (\textbf{D1}). During this process, users may interact with the overview page menu or incorrect pages, such as advertisement or deposit package pages. FetchAid sets the target button as "Fetch Package" on the overview page, while reporting tap errors (\textbf{D2}) and directing users to the "Back" button to recover from incorrect pages (\textbf{D3}).

We implement the following policy to determine the user's current page and target button:
\textbf{(1)} If both "Fetch Package" and "Deposit Package" buttons are detected, the user is on the overview page, and the target button is "Fetch Package."
\textbf{(2)} If the user successfully taps "Fetch Package," the QR code page appears. If its QR code banner (as seen in Fig. \ref{fig:overview}) is detected in 8 consecutive frames, a voice command guides the user to hold their phone steady for the QR code scan (\textbf{D2}). Meanwhile, FetchAid automatically prompts WeChat's built-in QR scanner to unlock the compartment (\textbf{D4}).
\textbf{(3)} If the user taps other buttons, the undesired page entry is detected. The "Back" button is detected 10 times in a row, and a voice command guides the user to return to the previous page (\textbf{D2}), setting the "Back" button as the target (\textbf{D3}).
\textbf{(4)} If the user successfully taps "Back" and returns to the overview page, the policy reverts to case \textbf{(1)} with voice feedback. In all cases, FetchAid provides voice feedback based on Fig. \ref{fig:touchingpoint} to guide the user.
\textbf{(5)} If the user's camera is not aimed at the touchscreen, the system guides them to the correct position. For example, if the touchscreen's bounding box appears at the top left of the phone screen, the user is prompted to "move to the top left" (\textbf{D7}).

Upon determining the target button and current page, we provide voice feedback based on the finger's and target button's locations to guide the user's finger movement. We define the finger's touching point (Point $A$) as the upper part of the finger pulp, typically used to interact with touchscreens. The object detection model recognizes the finger by learning a bounding box from the fingertip to the middle knuckle (Polygon $F$), as shown in the left half of Fig. \ref{fig:touchingpoint}. We track the center of the index fingertip, heuristically determined as $\frac{1}{3}$ upper part of Polygon $F$. We also track the geometric center (Point $B$) of the target button's learned boundary (Polygon $R$). By evaluating the relative position between $A$ and $B$, we generate voice feedback to guide the user in tapping the target button (\textbf{D1}), as illustrated in the right half of Fig. \ref{fig:touchingpoint}.

\subsection{Open Door Searching Phase}
Following the Touchscreen Interaction Phase, FetchAid guides the user's hand to the open compartment door (\textbf{D5}), necessitating tracking of the door's location relative to the phone's real-time position. After scanning the code, the user starts at the touchscreen, holding their phone towards it. FetchAid sets the phone's location as the origin and the determined location of the open compartment door as the goal. Navigation guidance not only directs the user in which way to move but also assists in avoiding potential collisions (\textbf{D6}).

\subsubsection{Obtaining Open Compartment's Location}
\begin{figure}[tbh!]
    \centering
    \includegraphics[width=\linewidth]{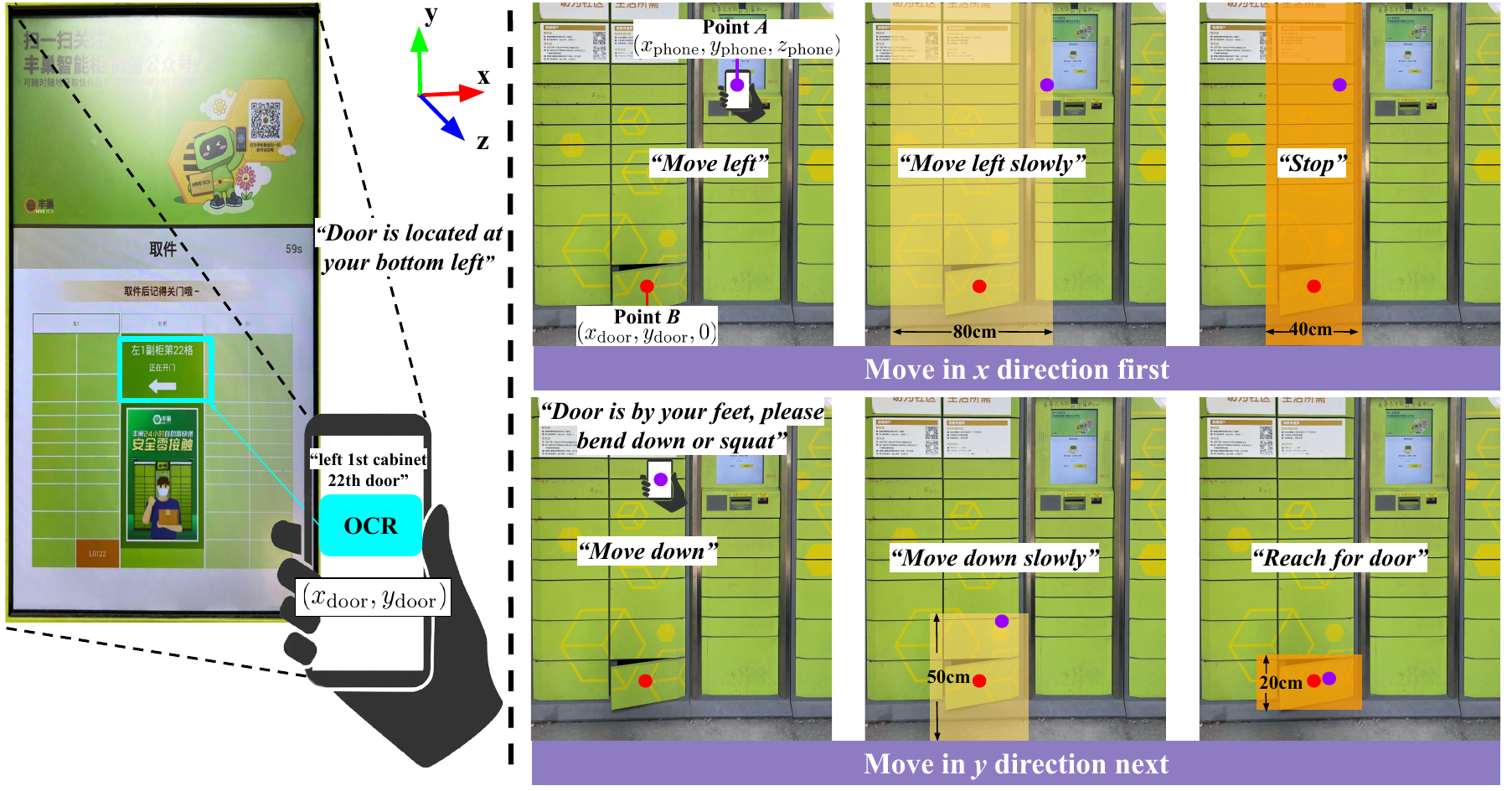}
    \caption{llustration of voice feedback in the Open Door Search Phase. \textbf{Left:} OCR reads the screen to derive the target compartment door's location (e.g., "Left 1st cabinet, 22nd door"). Knowing the user's average starting height and each door's dimensions, we accurately derive the target door location (Point $B$) and report its general direction (e.g., "bottom left"). \textbf{Right:} Point $A$ represents the user's mobile device center. Voice feedback is based on the relative location between the mobile phone (Point $A$) and the target door (Point $B$). Extra reminder voice feedback is provided if the door is too high or low (e.g., "Door is by your feet"). Voice feedback is automatically triggered based on user's motion.}
    \label{fig:opendoor}
    \Description{Figure 7 illustrates voice feedback in the Open Door Search Phase. The figure on the left shows OCR reads the screen to derive the target compartment door's location (e.g., "Left 1st cabinet, 22nd door"). Knowing the user's average starting height and each door's dimensions, we accurately derive the target door location (Point $B$) and report its general direction (e.g., "bottom left"). The figure on the right shows how the user follows FetchAid's guidance to the target door. Point $A$ represents the user's mobile device center. Voice feedback is based on the relative location between the mobile phone (Point $A$) and the target door (Point $B$). Extra reminder voice feedback is provided if the door is too high or low (e.g., "Door is by your feet"). Voice feedback is automatically triggered based on user's motion.}
\end{figure}

When the door opens, text appears on the touchscreen, providing the door's location with coordinates for the cabinet and compartment numbers. FetchAid implemented optical character recognition (OCR) with Apple Developer's live text streaming (data scanner) API~\footnote{\url{https://developer.apple.com/documentation/visionkit/datascannerviewcontroller}}. The data scanner detects text in real-time when the iPhone camera is on, and we use regular expressions to identify compartment location information. Given that parcel lockers have standardized dimensions, FetchAid can accurately derive the door's $xy$ position in world coordinates ($x_\text{door}, y_\text{door}$) relative to the origin, as depicted in the left part of Fig. \ref{fig:opendoor}. Simultaneously, FetchAid communicates the general direction of the open door based on the derived position, guiding the user towards it.


\subsubsection{Defining and Tracking User's Location} 
\label{sec: Defining and Tracking User's Location}
The default user origin is set at the touchscreen's center at the fourth compartment level, which corresponds to a 170cm-tall user holding their phone in front of their face or chest. To accommodate users of different heights, FetchAid allows height input and adjusts $y_\text{door}$ accordingly. To guide users, real-time location tracking using AR is essential. FetchAid employs a VIO system \cite{forster2016manifold, zhang2020dex} to track the device's coordinates ($x_\text{phone}, y_\text{phone}, z_\text{phone}$) robustly. The app leverages iOS' native ARKit toolkit to detect the parcel locker's facade as the plane anchor for the AR Session. ARKit's HitTest then tracks the transformation between the device and the plane anchor's center, determining the coordinates.

\subsubsection{Guiding the User to Find the Open Door} 
Navigating a large parcel locker can be physically demanding, so FetchAid aims to minimize frequent direction changes. Users are first guided horizontally along the $x$-axis, then vertically to the target $y$-range. Users are instructed to bend down, squat, or reach up as necessary. To prevent collisions, buffer ranges ($x_\text{door} \pm 40cm$) and ($y_\text{door} \pm 25cm$) are established, triggering a slow movement advice. Once within 10cm of the target, users are told to stop and reach slowly toward the open door. During this phase, speech feedback is augmented with vibration for more intuitive guidance, with vibration frequency increasing as the phone nears the target door. To prevent collisions, FetchAid also tracks the user's distance from the parcel locker's facade ($z_\text{phone}$) and reminds them to step back if closer than 30cm (\textbf{D6}).



\subsection{Support VoiceOver and Voice Control}
Since FetchAid is a smartphone application, we ensure its usable interface by supporting the phone screen reader (VoiceOver in iOS \cite{voiceover}) and spoken commands (Voice Control in iOS~\footnote{\url{https://support.apple.com/en-ca/HT210417}}). VoiceOver, popular among BLV iPhone users, reads or describes selected elements. Research has shown VoiceControl enhances Apple device accessibility \cite{voiceover}. To ensure VoiceOver compatibility, accessibility labels, and hints are added to all UI elements~\footnote{\url{https://developer.apple.com/documentation/accessibility/supporting_voiceover_in_your_app}}. Custom spoken commands are also available for BLV less familiar with screen readers, allowing them to use FetchAid by saying, "I want to fetch a package."

\subsection{Deployment}
We developed FetchAid using XCode 14 Beta on MacOS 12.5 and Swift language, ensuring optimal support for each software module. We trained EfficientNet-Lite2 with Python 3.8 and TensorFlow 2.8, exporting it to TensorFlow-Lite format. FetchAid is deployable on iOS devices compatible with iOS 16 \footnote{\url{https://support.apple.com/en-us/HT213411}}; it is self-contained, and requires no external resources. Utilizing a mobile-CPU-friendly deep neural network and Apple's native ARKit, the system is optimized for computational resources, enabling real-time operation and accurate voice feedback for users.

%% file: Doc/5-technical.tex
\section{Technical Evaluation}
We performed technical evaluations of the object detection module, OCR-compartment location detection module, and AR navigation module of the system to examine the robustness of our algorithms and models.

\subsection{Object Detection Model}
As discussed in Section~\ref{sec: Object Framing in Help with Camera Aiming}, previous blind photography assistive tools use computer vision on snapshots, which can result in motion blur and out-of-focus issues for hand-held mobile devices. FetchAid avoids this problem by not requiring users to take photos for object recognition. Instead, it processes a continuous stream of frames (inference time: 100 ms) and outputs results for later decision-making. Even if there is a single-frame object detection failure or error, it won't impact the model's accuracy during one inference session. Thus, we measure object detection model accuracy per session, not per frame. To validate our design choice, we evaluate the object detection model trained on data with and without augmentation. 

we assessed FetchAid's object detection performance through 80 trials, split evenly between models with and without data augmentation. Each trial encompassed all touchscreen interactions for package retrieval, and screen recordings were analyzed to determine object detection accuracy. We observed that detection accuracy without augmentation is 74.6\%, while with augmentation, it rises to 97.7\%. Data augmentation significantly improves accuracy for all object categories, except for the touchscreen, which is already accurate. This highlights the importance of data augmentation in boosting the object detection model's robustness. See Table \ref{table: obj-detection} in Appendix \ref{appendix} for detection accuracy for each object category.



\subsection{OCR-Compartment Location Detection}
\label{sec: OCR-Compartment Location Detection}
To evaluate OCR performance, we recorded the number of attempts until success, success rate, and recognition time at five camera angles (0\textdegree, 15\textdegree, 30\textdegree, 45\textdegree, and 60\textdegree) and three distance tiers (30-40, 40-50, and 50-60cm). Such evaluation condition setting is designed to adapt to varying forearm lengths and hand-holding preferences. For each angle and distance, we conducted ten tests, deeming more than three unsuccessful attempts a failure. The results demonstrated high accuracy even at steep angles, attesting to the robustness of our OCR. Refer to Appendix \ref{appendix}, Table \ref{table: ocr}, for a detailed breakdown of average OCR performance by angle and distance.

We assessed OCR module reliability by comparing the derived location to the ground truth. We chose two parcel lockers (single and two-column main cabinets) and set up 30 target compartment locations for each locker. For each target, we measured the distance from its center to the touchscreen center on the $x$ and $y$ axes ($d_x$, $d_y$) as the ground truth. After OCR detects door coordinates, our algorithm calculates the target distances on the $x$ and $y$ axes ($d_x'$, $d_y'$). We then compute the differences in $x$ ($|d_x \ - d_x'|$, $M= \ 1.6$, $SD \ = 1.2$) and $y$ ($|d_y \ - d_y'|$, $M= \ 0.8$, $SD \ = 1.4$) axes. Both mean values are less than 2 cm, indicating that the OCR-based door location derivation is highly accurate.

\subsection{AR Navigation}
We test the AR Navigation module's accuracy using the same two parcel lockers and 30 locations for each cabinet as in section~\ref{sec: OCR-Compartment Location Detection}. We mark the phone's location ($l$) when the voice feedback "Reach for the door" is triggered and measure the distance from $l$ to the center of the targeted compartment ($M \ = 10.1$, $SD \ = 3.8$). The average and standard deviation of the measurements are shown in Table~\ref{table: ar} in Appendix \ref{appendix}. As explained in Section~\ref{sec: Defining and Tracking User's Location}, given the width and height of each compartment, an error of around 10 cm is considered within the safe range for accurately touching the compartment. Thus, AR-based navigation is a reliable technique in the overall process.


%% file: Doc/6-userstudy.tex
\section{User Study}
Our user study aimed to evaluate the impact of FetchAid on enabling BLV users to fetch packages from typically inaccessible parcel lockers. Based on the data collected, our quantitative evaluations and analyses focused on the success rate and completion time of FetchAid's two stages, while our qualitative assessments examined the effort and frustration involved in the package fetching process. The study is approved by the university ethics review board.  

\begin{figure}[tbh!]
    \centering
    \includegraphics[width=\linewidth]{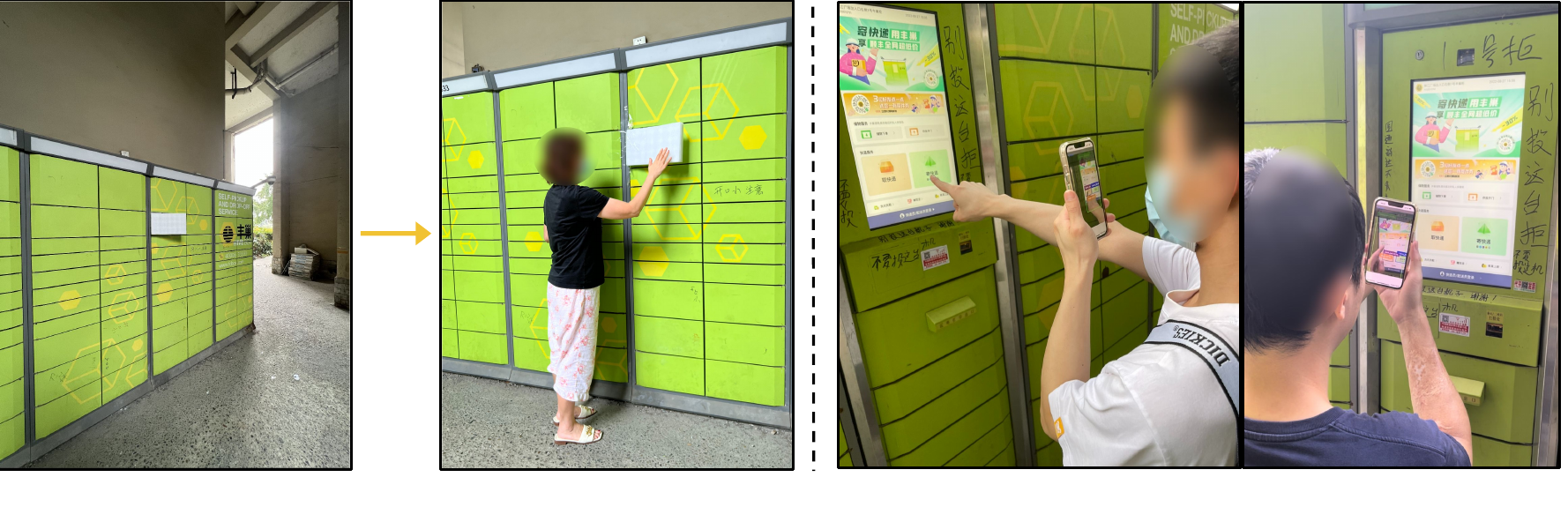}
    \caption{\textbf{Left: }We used polymeric foam to simulate the open door, which users needed to touch during our user study. Additionally, we replicated the "click" sound of an unlocking door by gently tapping the target door with a stick. \textbf{Right: }Screenshots from the user study recordings show BLV participants interacting with the parcel locker interface using FetchAid.}
    \label{fig:poly}
    \Description{Figure 8 left illustrates the simulated open door made from polymeric foam. The right image depicts a user touching the simulated door during our study. We also emulated the "click" sound of an unlocking door by gently tapping the target door with a stick. The right image includes screenshots from recorded user study videos, showcasing BLV participants interacting with the parcel locker interface using FetchAid.}
\end{figure}

\subsection{Participants}
\st{We recruited 12 BLV participants with varying conditions of visual impairment for the full user study. We recruited participants with different experience levels of smartphone use and of parcel locker use in order to see how FetchAid might be able to assist users with varying understanding of the package fetching experience using parcel lockers. The participants' demographic information is shown in Table \ref{tab:user-info}.}

\rv{
The study recruited 12 BLV participants with varying conditions of visual impairment (The participants' demographic information is shown in Table \ref{tab:user-info}). There were three constraints in sampling. First, they had to live in a specific city in China and be willing to travel to the user study field site. Second, they had to be people with blind and low vision. Third, they are all smartphone users. Due to these constraints, we initially recruited five participants through a disability online support group local to the specific city. Subsequently, the snowball sampling technique was employed, where we asked these participants to forward user study invitations to their contacts that satisfy the requirements.
}

The recruited group has different experience levels of smartphone use and of parcel locker use. This is helpful for the researchers to understand how FetchAid can assist users with varying technology familiarity and package fetching experience using parcel lockers.

\subsection{Apparatus}
In our user study, we installed FetchAid on an iPhone 12 Pro Max running iOS 16 Beta, which we provided to participants during the study. The object detection system was trained using the previously mentioned manually labeled and augmented dataset, without test-time fine-tuning. We conducted the study on five different, actively used public parcel lockers with varying sizes (i.e. consisting of different numbers of cabinets) and in different locations.

\input{supplements/user_info.tex}

\subsection{Setup and Experimental Design}
The user study was conducted using six different HiveBox parcel lockers, all without accessibility features, as these are the commonly used lockers in the area where the researchers and participants live. We split the user evaluation into two tasks, \textit{Touchscreen Interaction} and \textit{Open Door Search}, to better assess FetchAid's usefulness in each phase. In \textit{Touchscreen Interaction}, participants used a different parcel locker for their each trial to prevent them from memorizing the "Fetch Package" button's location on the touchscreen. The target button locations varied due to differing touchscreen installation heights and UI, introducing more randomness to this task. For \textit{Open Door Search}, we avoided memorization by changing the target compartment door position and parcel locker in each trial. To ensure a fair comparison between baseline and experimental trials, we maintained consistent distances from the participant's starting location to the target door location in each trial pair.

Since parcel locker systems require an actual courier delivery to open a compartment door, we taped polymeric foam to the target door to simulate an open door, as shown in Fig. \ref{fig:poly}. We taped compartment door locations to the touchscreen so the OCR system could detect the text in the Open Door Search task without a real package. In each trial, we simulated the "click" sound of an unlocked compartment door by tapping the target door with a stick as an auditory cue for the participant.

For each task, we conducted paired testing with two conditions: (i) baseline - where the participant performs the task independently on a HiveBox; and (ii) experimental - where they used FetchAid on the provided iPhone. We employed a within-subjects design and counterbalanced the condition order, with 50\% of users completing a task without FetchAid first. Each task and condition required four trials, limited to 60 seconds per trial. The study setup is depicted in Fig. \ref{fig:poly}.


\subsection{\rv{Data Collection}\st{Procedures}}
\rv{We offered ride-sharing for participants to travel from home or workplace to the site.} After participants arrived, they were introduced to the study and signed consent forms. \rv{The study began with a} \st{They began} training phase to familiarize themselves with the parcel locker interfaces and FetchAid. We demonstrated the "Fetch Package" button's position on the touchscreen and the simulated open door, essential for those unfamiliar with parcel lockers. 

Once the training was complete, participants performed the two tasks in their assigned order. In the baseline trials for \textit{Touchscreen Interaction}, we informed the participant if they tapped the wrong button or successfully revealed the QR code. We recorded the time taken to scan the QR code. For an \textit{Open Door Search} trial, we recorded the completion time when the participant reached the foam door. \rv{Upon completion of both tasks, participants were asked to rate their overall experience in collecting packages from parcel lockers, both with and without the FetchAid system, based on the NASA Task Load Index (TLX) questionnaire (\cite{hart2006nasa}). To gain a deeper understanding of these ratings, the researcher then conducted semi-structured interviews. These interviews focused on gauging the context behind the participants' feedback, allowing for a more comprehensive analysis of their experiences.}

Each participant spent approximately 30 minutes in the study and was compensated according to local standards. The entire study was video and audio recorded for analysis. \rv{The study's internal validity is supported by triangulation, using experiment data, task load ratings, and interviews. These diverse methods enabled a thorough assessment of FetchAid's impact on package retrieval. Participant ratings were cross-verified with interview insights, and both were compared with experimental outcomes.This approach enriches understanding by using qualitative insights to contextualize quantitative data, revealing deeper reasons behind FetchAid's impact.}

\rv{

\subsection{Data Analysis}

\forceindent \textbf{Task Data} For each task, we conducted two-tailed paired t-tests ($\alpha = 0.05$) to compare the success rate and completion time (only recorded for successful trials) between the baseline and experimental trials. 
Specifically, for each participant, we compared the success rates and completion times averaged across 4 trials for both baseline and experimental conditions. See Table \ref{tab:user-result} in \ref{appendix} for the detailed data.

\textbf{TLX Questionnaire} It gathered subjective feedback on mental, physical, and temporal demands, performance, effort, and frustration. We analyzed the results of Wilcoxon signed-rank tests ($\alpha = 0.05$) on participants' ratings (Fig. \ref{fig:ratings}).

\textbf{Interviews} The interviews were transcribed post-study and analyzed through content analysis to establish empirical links to thematic findings. Initial categorization focused on responses directly relevant to task components, followed by thematic coding. This coding was tailored to reflect actual challenges in the package retrieval task, aiming to deepen our understanding of user experience.

}


%% file: supplements/user_info.tex
\begin{table}[tbh!]
\centering
\resizebox{\linewidth}{!}{
\begin{tabular}{|l|l|l|l|l|l|l|}
\hline
\textbf{ID} & \textbf{Gender} & \textbf{Age Range} & \textbf{Occupation} & \textbf{Vision} & \textbf{Smartphone Use} & \textbf{Parcel Locker Use} \\ \hline
P1 & Male & \multicolumn{1}{l|}{20-30} & Student & Light, Color & iPhone, \textgreater{}3 years & Once a week \\
P2 & Female & \multicolumn{1}{l|}{20-30} & Musical performer & Light & iPhone, \textgreater{}3 years & Once a week \\
P3 & Male & \multicolumn{1}{l|}{20-30} & Freelancer & Light & iPhone, \textgreater{}3 years & Seldom \\
P4 & Male & \multicolumn{1}{l|}{30-40} & Career Consultant & Light & Android, 1-2 years & Once a month \\
P5 & Female & \multicolumn{1}{l|}{20-30} & Musical performer & Light, Color & iPhone, 1-2 years & Once a month \\
P6 & Male & \multicolumn{1}{l|}{20-30} & Student & Light, Color & Android, 1-2 years & Once a week \\
P7 & Male & \multicolumn{1}{l|}{20-30} & Massager & Light & Android, \textgreater{}3 years & Once a month \\
P8 & Male & \multicolumn{1}{l|}{40-50} & Massager & Blind & iPhone, \textgreater{}3 years & Never \\
P9 & Female & \multicolumn{1}{l|}{50+} & Retired & Blind & iPhone, 1-2 years & Never \\
P10 & Male & \multicolumn{1}{l|}{20-30} & Student & Light, Color & iPhone, \textgreater{}3 years & Twice a month \\
P11 & Male & \multicolumn{1}{l|}{20-30} & Freelancer & Light, Color & Android, 1-2 years & Never \\
P12 & Female & \multicolumn{1}{l|}{40-50} & Retired & Light, Color & iPhone, \textless{}1 year & Never \\ \hline
\end{tabular}
}
\caption{\label{tab:user-info} Participants' information from the user study. Each participant has different visual impairment conditions of different severity. In the Vision column, ``Light'' represents the inability to perceive light changes, ``Light, Color'' represents the inability to perceive neither light nor color changes, and ``Blind'' represents completely no vision. }
\end{table}

%% file: Doc/7-results.tex
\section{Results}
\label{sec: results}
For each task, we conducted two-tailed paired t-tests ($\alpha = 0.05$) to compare the success rate and completion time (only recorded for successful trials) between the baseline and experimental trials. 
Specifically, for each participant, we compared the success rates and completion times averaged across 4 trials for both baseline and experimental conditions. See Table \ref{tab:user-result} in \ref{appendix} for the detailed data.

\subsection{Touchscreen Interaction Task}

Results show that using our system significantly improves the success rate compared to not using it, $t(11) = -2.930, p = 0.014$. In 15 trials, participants missed the "Fetch Package" button and tapped the wrong button. However, in 14 of these cases, they corrected their errors by listening to FetchAid's voice feedback. The greatest improvement was seen among inexperienced parcel locker users (P3, P8, P9, P11, and P12), with a 50\% increase in success rates. This is likely because real-time voice feedback eliminated the need for familiarity with the parcel locker's interface to successfully tap the "Fetch" button. Participants appreciated voice feedback: \textit{"I was daunted by the complexity of the parcel locker's interface at first during baseline trials, but the app's voice feedback really helped me a lot because I just need to follow what it says to me." - P8.}  When comparing only completed trials, no significant difference in completion time was found between the baseline and experimental tasks, $t(9) = -1.279, p = 0.233$. Very experienced participants stated they \textit{"...tapped the button with muscle memory" - P6} in the baseline condition, which took less time than the finger detection process. However, they spent less time launching the QR code scanner on the phone: \textit{"The new app handles QR code scanning itself, and that saves a lot of time for me even though I know how to navigate the parcel locker's interface." - P1}


\subsection{Open Door Search Task}
The task success rate significantly increased with FetchAid, $t(11) = -2.803, p = 0.017$, and completion time significantly decreased, $t(11) = -2.423, p = 0.034$. Without our system, participants often spent considerable time locating the open door by touching and feeling the parcel locker: \textit{"I pay extra attention to the 'click' noise from the compartment door when it unlocks and try to find the door by touching the cabinet and moving toward the noise. Sometimes I miss the door and fear being poked by the cabinet parts." - P9}. Voice feedback informs users where to expect the open compartment door and provides real-time guidance, eliminating the need for constant touch and reducing both time and stress from the fear of being poked.

\subsection{Error Recovery}
Section \ref{sec: results} demonstrated that FetchAid significantly improved participants' success rates, with most reaching or nearly reaching 100\%. Despite these high success rates, we also analyzed the few failure cases. In our experiments, we identified ten instances of erroneous clicks and recovery attempts. Seven of these occurred when participants clicked a button adjacent to the "fetch package" button (e.g., "deliver package" or "storage order"). In these cases, participants were redirected to a distraction page, but FetchAid's error recovery support helped them return to the previous page in five out of the seven cases (71.4\%). In the remaining two cases, participants failed to click the "back" button within the allotted time. In the other three cases, voice instructions were correct, but participants did not click any buttons on the screen.

The data reveals that most errors were successfully recovered, and in unsuccessful cases, time limitations were the primary issue. As each participant has varying finger dexterity and reaction abilities, allowing more time could have resulted in a successful recovery. Furthermore, we observed that most errors occurred during participants' initial attempts, and the probability of errors or recovery decreased with subsequent attempts, suggesting that the success rate can improve with practice.

\subsection{Subjective Feedback}
\begin{figure}[tbh!]
    \centering
    \includegraphics[width=\linewidth]{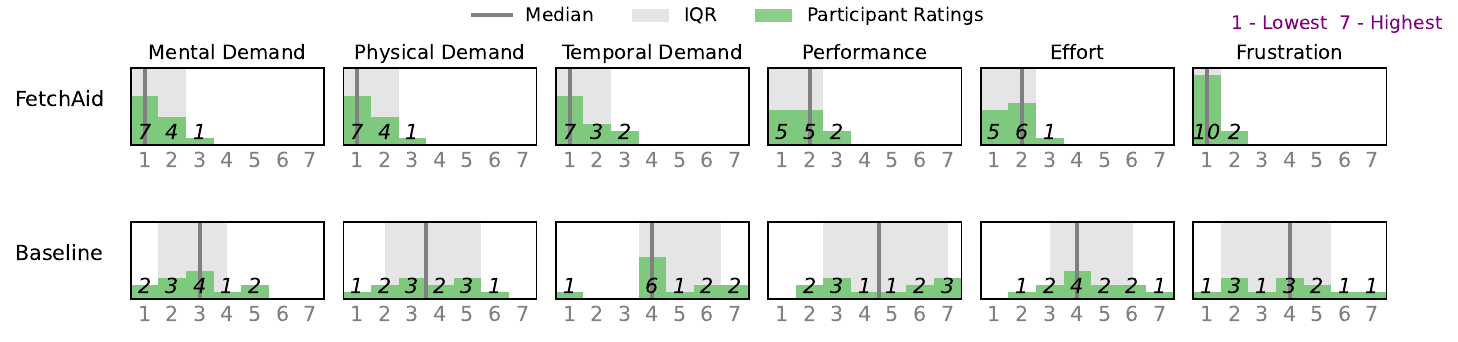}
    \caption{Participants' questionnaire ratings on a scale of 1-7 using NASA Task Load Index. \textbf{Top: }ratings for package fetching with FetchAid. \textbf{Bottom: }ratings for package fetching without FetchAid. For the performance scale, a lower rating means a better performance.}
    \label{fig:ratings}
    \Description{Figure 9 shows participants' NASA Task Load Index questionnaire ratings on a scale of 1-7. The first row is ratings for package fetching with FetchAid in histograms. The second row is ratings for package feting without FetchAid. For performance scale, a lower rating mean a better performance. For every rating for FetchAid, the medians and IQR are more left-skewed than the baseline.}
\end{figure}

\st{Participants were asked to rate their \textit{overall} experience of collecting packages from parcel lockers with and without FetchAid. The questionnaire gathered subjective feedback on mental, physical, and temporal demands, performance, effort, and frustration. We analyzed the results of Wilcoxon signed-rank tests ($\alpha = 0.05$) on participants' task load ratings (Fig. \ref{fig:ratings}) and their interview responses.}

\rv{The analysis on TLX ratings and interviews showed that FetchAid significantly reduced BLV users' frustrations with parcel lockers, which stemmed from the touchscreen's complexity, lack of response, and difficulty distinguishing the unlocked door from other compartments \cite{chi22blvPackage}.} \st{A recent study identified BLV users' primary frustrations with parcel lockers as the touchscreen's complexity, lack of response, and difficulty distinguishing the unlocked door from other compartments \cite{chi22blvPackage}. } FetchAid offers an accessible alternative interface, significantly reducing frustration and improving performance when using parcel lockers (Fig. \ref{fig:ratings}). The assistive application provides regular real-time feedback, minimizing aimless exploration for BLV users. For example, P11 appreciated the Open Door Search guidance, \textit{``The Open Door Search guidance is much more reliable, especially its exact measurement... Usually, I capture a clicking sound to locate the open door, but sometimes I get distracted and lose it in the noisy surroundings. Then I will have to traverse the entire cabinet on one side.'' - P11} Another participant \textit{P3} shared a similar thought on the touch screen interaction, \textit{``I am never sure what function button I have tapped on the touch screen, so I just use my phone to scan the QR code away, and the Voiceover will tell me if I have tapped the right Fetch Package button or not... The app assured me a lot more, knowing that it will correct me even when I tap wrong.''} 

Retrieving a package from a parcel locker involves multiple steps, requiring additional effort from BLV users, such as knowing when to open the QR scanner and coordinating with the correct page selection after tapping a button. FetchAid reduces this effort, subsequently decreasing the stress of the retrieval process. Participants reported lower mental, physical, and temporal demands and overall effort when using FetchAid compared to baseline trials without the system (Fig. \ref{fig:ratings}). The intuitive real-time voice feedback ensures users only need to follow instructions, making the process \textit{``... almost a no-brainer... I don't need to worry about if I clicked the wrong button or what to do next because the app will give me voice instructions.''- P1}

FetchAid also helps reduce errors that are typically challenging for BLV users to address \cite{chi22blvPackage}. Participants appreciated the automatic QR code scanning, which replaced the manual entry of the pickup code, preventing mistakes due to unfamiliar keypad layouts (P7). Additionally, FetchAid helps users avoid issues like collisions and accidental door closures during navigation (P5).

%% file: Doc/8-discussion.tex
\section{Discussion}
FetchAid allows BLV users  to retrieve packages from parcel lockers efficiently, offering error-recovery and real-time, context-aware audio feedback for touchscreen use and safe compartment navigation. We will explore extending FetchAid for different user interfaces, supporting various parcel lockers, and adapting to other public touchscreen devices.Finally, we will discuss limitations and potential future work.

\subsection{Novelty of FetchAid: A Tailored Solution for Parcel Locker Interaction}
FetchAid stands apart from existing smartphone apps for vision detection, as it specifically addresses the challenges BLV users face when fetching packages from parcel lockers. For instance, While Seeing AI, a popular general-purpose vision detection app \footnote{\url{https://www.microsoft.com/en-us/ai/seeing-ai}}, excels in identifying key objects like products and people, it falls short in tracking touchscreen elements or providing accurate guidance to the specific locker door. Vizlens, which relies on cloud workers and targets static appliance interfaces \cite{guo2016vizlens}, is not cost-effective or suitable for dynamic parcel locker touchscreens.

By focusing on this particular task, FetchAid offers a tailored user experience that accommodates the unique requirements of parcel locker interactions. Its specialized design and functionality ensure that BLV users can effectively navigate and interact with parcel lockers, providing a much-needed solution for the special challenge with touchscreen-based large machines.

\subsection{Error Recovery Support}
FetchAid's error-recovery mechanism improves success rates and reduces frustration, affirming the hypothesis that BLV users are more likely to succeed and persist when assisted during task failures \cite{vigo2013}. This support allows users to correct mistakes by detecting incorrect entries, providing notifications, and recognizing the "Back" button. 

Users are given cues if there were multiple wrong entries, but some errors remain irrecoverable due to additional incorrect inputs and time-consuming recovery. Video analysis indicates that early impatient actions, like tapping ahead of instructions or moving too quickly for the audio feedback, can lead to consequential errors. This supports the idea that lengthy responses can cause user frustration and impatience \cite{Branham2019}. Users typically adjusted and succeeded in later trials. Future developments should include personalization to match feedback timing with individual user pacing. While our system doesn't automatically reopen closed doors, it can guide users back to restart the process with FetchAid.

\rv{\subsection{Environmental Cues}
We acknowledge that the simulated open door scenario in our study may not fully replicate real-world cues like the distinct sound of a door opening. Despite this, the successful completion of our experimental tasks indicates that the system was effective in supporting users even without these specific cues. This finding is consistent with previous research~\cite{chi22blvPackage}, which underscores the unreliability and limited effectiveness of external environmental cues, such as the audio cues generated when a compartment is open. In this study, BLV participants noted the inconsistency of audio cues, which varied with the degree of door opening. They often resorted to physically holding the surface of the parcel locker and detecting the open door by touch. Moreover, it is common for parcel lockers to be in outdoors, so the surrounding noise may also interfere with the audio cues of door opening. Nevertheless, future work could investigate reliable ways of integrating audio cues with our proposed approach to better assist BLV people. 
}

\subsection{Supporting Potential User Interface Updates}
Current data augmentation techniques like flipping, rotation, and random color adjustment add robustness to the model, ensuring it adapts to minor changes in the UI's appearance and location~\cite{narayanan2020deep}. For major UI updates, such as new arrangements, researchers could collect and annotate new data (around 50 snapshots with updated UI plus augmentation), fine-tune the object detection network, and redeploy it. While FetchAid presently relies on researchers for data collection following UI updates, our plan is to open-source the data collection platform once the FetchAid is launched. This strategy empowers users to upload and label new interfaces, significantly diminishing researcher intervention and keeping FetchAid current with machine updates.


\subsection{Generalization to Other Machines} 
FetchAid was designed for HiveBox parcel lockers and may not directly apply to parcel lockers from other companies. Yet, as Fig. \ref{fig:worldwideparcel} indicates, parcel lockers globally have similar features and user operations. For instance, DHL Packstations require selecting "Pick Up" on the touchscreen and scanning a pickup code \footnote{\url{https://www.dhl.de/en/privatkunden/pakete-empfangen/an-einem-abholort-empfangen/packstation/bedienung-packstation.html}}. With minimal modification, FetchAid's design and interaction flow can adapt to various lockers by customizing button labels and updating dimensions.

The design logic and key components of FetchAid could aid accessibility research for other public touchscreen devices and spatial navigation tasks. Its interaction paths and error recovery are applicable to similar touchscreen appliances, and its AR navigation is valuable for accurate movement in confined spaces, suggesting adaptability to diverse products with slight logic adjustments.

\begin{figure*}[tbh!]
    \centering
    \includegraphics[width=\textwidth]{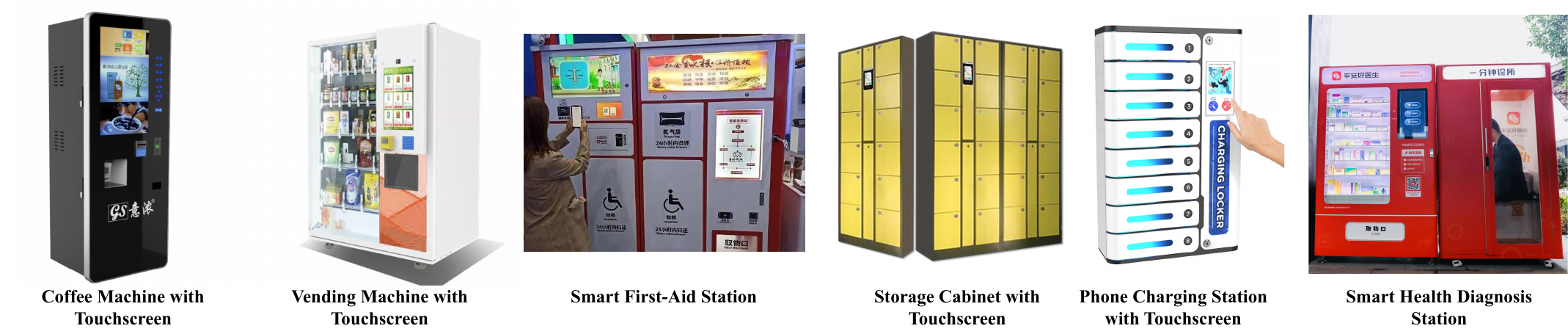}
    \caption{Products that FetchAid could potentially be generalized to. FetchAid can be generalized to products that would require touchscreen interaction and navigation needs. To extend the use cases, FetchAid needs more data collection to train the object detection model for the touchscreen interaction module as well as a task-specific logic for target position calculation. }
    \label{fig:extension}
    \Description{Figure 10 shows the products that FetchAid could potentially be generalized to, which are: coffee machine with touchscreen, vending machine with touchscreen, smart first-aid station, storage cabinet with touchscreen, phone charging station with touchscreen, and smart health diagnosis station. To summarize, FetchAid can be generalized to products that would require touchscreen interaction and navigation needs.}
\end{figure*}

Adapting FetchAid for other machines involves three steps: collecting UI images for object detection model retraining, modifying the error-recovery logic, and updating the Open Door Search Module’s target position. Fig. \ref{fig:extension} showcases a range of products that could potentially benefit from FetchAid's application, all of which are touchscreen-based and lack accessibility features. AR navigation could be particularly beneficial in confined spaces, like first-aid or phone charging stations. To encourage further accessibility enhancements, we aim to open-source our code, inviting the community to build upon FetchAid and enhance the accessibility of parcel lockers globally.

\subsection{Combination of FetchAid with Other Assistive Technology for BLV}

FetchAid starts with the assumption that visually impaired users are already near the parcel locker's touchscreen, but they also need to navigate from their initial location, like home, to the locker. Although FetchAid doesn't currently offer this broader navigation assistance, the challenge is well-explored in existing research. NavCog3, for instance, offers detailed indoor navigation guidance \cite{sato2017navcog3}, and recent AR smartphone apps provide outdoor obstacle detection \cite{coughlan2017ar4vi, kaul2021mobile}. In principle, FetchAid could integrate with these apps, providing door-to-door assistance, from home to locker and back. To support such integration, we plan to open-source our code.

\subsection{Limitation and Future work}
\forceindent \textbf{Robustness}. 
FetchAid sometimes struggles with finger and graphical element detection on touchscreens in low-light conditions, leading to first-task timeouts due to the object detection network failing to recognize the user's finger, resulting in incorrect directions. Integrating low-light image enhancement techniques could improve detection accuracy \cite{CVPRLowLightMa, CVPRLowLightFeng}. Additionally, in the second phase, rapid user rotation can hinder AR plane anchor detection, causing delayed voice feedback. These issues stem from the hardware and VIO algorithms used. Future improvements could include a hybrid ML-based plane detector in mobile AR for consistent plane detection in varied environments \cite{planedetect}.Another area for enhancement is handling extreme failure scenarios. While our study didn't experience accidental door closures, this risk should be addressed. Future developments could involve an error recovery mechanism for the second phase, allowing FetchAid to detect and correct deviations from the intended path or guide users back to the touchscreen for a restart if needed.

\textbf{Voice Feedback Design}.
Our findings indicate that participants required fewer or different voice instructions over time. For example, users sometimes missed the end due to the failure to adjust their speed of movement, which relates to the lack of more detailed guidance. An interesting future research question is how to provide more effective and personalized voice instructions or audio feedback. Indeed, prior work has studied the effectiveness of different voice and audio feedback for BLV. One possible direction might be to develop an adaptive interface that allows BLV to customize voice feedback based on their preferences \cite{AudFeedback}.

\st{\textbf{Long Term Study for Error Recovery Design}.
Our error recovery mechanism promptly detects errors made by BLV, such as accidentally clicking on an advertisement. While participants could recover from such mistakes, other error types may prove more challenging. To better understand BLV's long-term parcel locker usage and the errors they encounter, it is important to collect comprehensive data over extended periods (e.g., months or even years). This dataset would enable the development of more sophisticated error-recovery mechanisms.}

\rv{\textbf{Longitudinal Study}. Our study aims to conduct a comprehensive longitudinal analysis of BLV users' interactions with parcel locker systems. This includes a long-term examination of error-recovery mechanisms, particularly on accidental clicks beyond advertisements. However, the scope extends beyond error recovery. We intend to observe and analyze how BLV users' engagement with the system evolves over extended periods (e.g., months or years). This will provide insights into long-term user behavior and adaptation strategies. }

\textbf{Design for Different Residual Vision of BLV}.
Our study suggested that participants' residual vision affects their usage behaviors. Blind participants tended to rely heavily on audio and haptic feedback, whereas those with low vision exhibited more autonomy, being confident in their residual vision. Blind participants benefited more in terms of success rate improvement. However, residual vision may deteriorate, leading to increased reliance on the app. A natural extension of this work involves adjusting the frequency of instructions to better accommodate users with residual vision and adapt to their pace when using parcel lockers.

\textbf{Other Smartphone Operating Systems (OS)}. Results in tables \ref{tab:user-result} (in Appendix \ref{appendix}) and \ref{tab:user-info} indicate no significant performance differences based on BLV's prior experience with smartphones or different smartphone operating systems. This suggests FetchAid is easy and intuitive to use without requiring familiarity with a specific smartphone OS. To further accessibility, we plan to open-source our code, enabling the development of Android versions of FetchAid for a broader range of smartphone users.

%% file: Doc/9-conclusion.tex
\section{CONCLUSION}
Parcel lockers have gained popularity in urban environments, yet their accessibility for BLV users remains limited. To address this issue, we introduce FetchAid, a stand-alone smartphone application that assists BLV users in using parcel lockers efficiently through intuitive, real-time voice feedback and guidance powered by a deep-learning vision system and an augmented reality system. Our evaluation of FetchAid with real BLV users and parcel lockers demonstrates its effectiveness and robustness. Data and feedback from participants show a 30\% increase in touchscreen interaction success rate and an 11\% improvement in open compartment door search success rate. Additionally, participants reported reduced frustration and effort while using FetchAid to retrieve packages, thanks to the real-time voice feedback.

As touchscreen devices become increasingly prevalent, FetchAid represents a significant contribution towards creating cost-effective, third-party assistive tools. These tools, which require no external apparatus or wearables but just smart devices, enhance touchscreen operations, error recovery, and object localization with safety-guiding mechanisms. FetchAid stands as a foundational step in the HCI community for developing more accessible interface interaction technologies. Leveraging efficient deep neural networks and augmented reality capabilities of smartphones, we can compensate for the inaccessibility of modern appliances and machines, thereby improving usability for BLV users.


%% file: Doc/appendix.tex
\section{Implementation Detail}
\begin{figure}[tbh!]
    \centering
    \includegraphics[width=\linewidth]{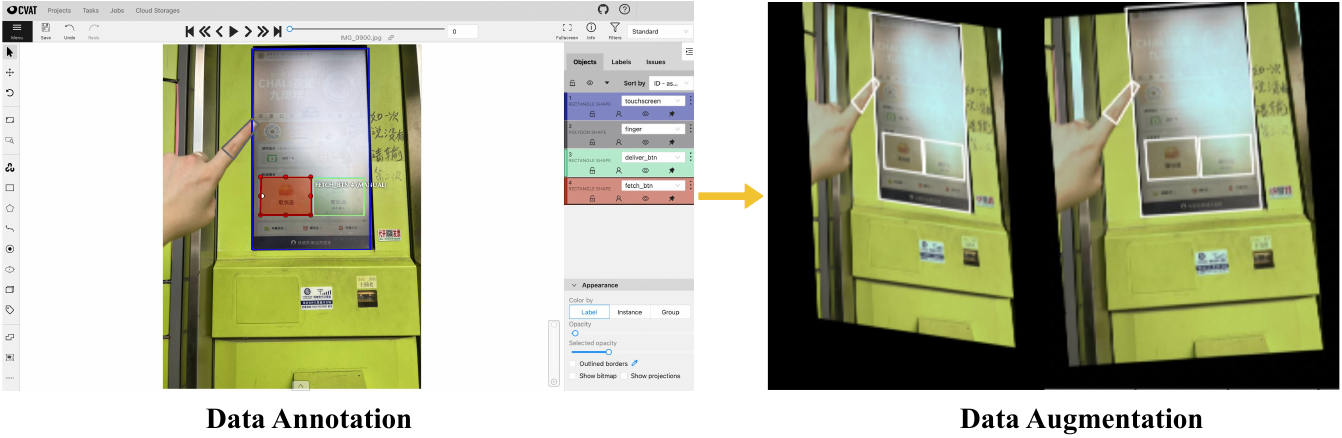}
    \caption{Illustration of the training data labeling and augmentation process. \textbf{Left: } CVAT data annotation user interface. The picture shown here has the right hand holding the phone and the left-hand index finger pointing. \textbf{Right: } Examples of augmented training data. When labeling the overview page, we label the page itself (purple), the fingertip (grey), the ``Fetch Package'' button (red), and the ``Deposit Package'' button (green).}
    \label{fig:cvat}
\end{figure}

\section{Technical Evaluation}
\label{appendix}

\begin{table}[tbh!]
\resizebox{\linewidth}{!}{
\begin{tabular}{|c|c|c|c|c|c|c|}
\hline
                     & Back Button & Deposit Button & Fetch Button & QR Code Page Banner & Finger & Touchscreen \\ \hline
Without Augmentation & 0.64        & 0.52           & 0.82         & 0.83                 & 0.69   & 0.98        \\ \hline
With Augmentation    & 1           & 0.94           & 1            & 1                    & 0.94   & 0.98        \\ \hline
\end{tabular}
}
\caption{Object Detection Model Accuracy with/without augmentation: success detection accuracy on each type of objects}
\label{table: obj-detection}
\end{table}


\begin{table}[tbh!]
\resizebox{\linewidth}{!}{
\begin{tabular}{|c|c|c|c|c|}
\hline
Camera-Plane   Angle & Camera-Plane Distance & Number of Attempts & Success Rate & Recognition Time (s) \\ \hline
0\textdegree                    & 30-40 cm                 & 1.2                & 1            & 2.6              \\ \hline
0\textdegree                       & 40-50 cm                  & 1.3                & 1            & 2.4              \\ \hline
0\textdegree                       & 50-60 cm                  & 1.1                & 1            & 2.4              \\ \hline
15\textdegree                      & 30-40 cm                  & 1                  & 1            & 2.5              \\ \hline
15\textdegree                      & 40-50 cm                  & 1.2                & 1            & 2.3              \\ \hline
15\textdegree                      & 50-60 cm                  & 1.5                & 1            & 2.8              \\ \hline
30\textdegree                      & 30-40 cm                  & 1.2                & 1            & 2.8              \\ \hline
30\textdegree                      & 40-50 cm                  & 1.1                & 1            & 1.5              \\ \hline
30\textdegree                      & 50-60 cm                  & 1.8                & 1            & 3.4              \\ \hline
45\textdegree                      & 30-40 cm                  & 1.3                & 1            & 2.4              \\ \hline
45\textdegree                      & 40-50 cm                  & 2                  & 0.9          & 3.6              \\ \hline
45\textdegree                      & 50-60 cm                  & 1.2                & 1            & 2.3              \\ \hline
60\textdegree                      & 30-40 cm                  & 1.5                & 1            & 3                \\ \hline
60\textdegree                      & 40-50 cm                  & 2.2                & 0.9          & 3.6              \\ \hline
60\textdegree                      & 50-60 cm                  & 1.2                & 1            & 2.5              \\ \hline
\end{tabular}
}
\caption{OCR Detection Test: Average performance under different angles of the camera to the touchscreen plane.}
\label{table: ocr}
\end{table}

\newlength\mylen
\begin{table}[tbh!]
\resizebox{\linewidth}{!}{
\begin{tabular}{|l|cc|cc|cl|}
\hline
 & \multicolumn{2}{c|}{OCR $x$-Dist. Error} & \multicolumn{2}{c|}{OCR $y$-Dist. Error} & \multicolumn{2}{l|}{Achieved Dist. Error} \\ \cline{2-7} 
 & \multicolumn{1}{c|}{AVG (cm)} & STD (cm) & \multicolumn{1}{c|}{AVG (cm)} & STD (cm) & \multicolumn{1}{c|}{AVG (cm)} & \multicolumn{1}{c|}{STD (cm)} \\ \hline
Standard Parcel Locker & \multicolumn{1}{c|}{1.53} & 1.23 & \multicolumn{1}{c|}{0.87} & 1.45 & \multicolumn{1}{c|}{10.9} & \multicolumn{1}{c|}{3.59} \\ \hline
Non-Standard Parcel Locker & \multicolumn{1}{c|}{1.67} & 1.16 & \multicolumn{1}{c|}{0.83} & 1.44 & \multicolumn{1}{c|}{9.3} & \multicolumn{1}{c|}{3.74} \\ \hline
\end{tabular}
}
\caption{AR Navigation Evaluation. We first report the errors of OCR-based location derivation with respect to ground-truth door location in both $x$ and $y$ directions. We also report the final achieved distance error as the L2 distance from the phone to the target door location.}
\label{table: ar}
\end{table}

\section{User Study Detail}
\begin{table}[H]
\resizebox{\linewidth}{!}{
\begin{tabular}{|l|llll|llll|}
\hline
\multicolumn{1}{|c|}{\textbf{}}          & \multicolumn{4}{c|}{\textbf{Touchscreen Interaction}}                                                               & \multicolumn{4}{c|}{\textbf{Open Door Search}}                                                                      \\ \cline{2-9} 
\multicolumn{1}{|c|}{\textbf{User   ID}} & \multicolumn{2}{c|}{Success Rate}                        & \multicolumn{2}{c|}{Completion Time (s)}                 & \multicolumn{2}{c|}{Success Rate}                        & \multicolumn{2}{c|}{Completion Time (s)}                 \\ \cline{2-9} 
\multicolumn{1}{|c|}{\textbf{}}          & \multicolumn{1}{c|}{Baseline} & \multicolumn{1}{c|}{App} & \multicolumn{1}{c|}{Baseline} & \multicolumn{1}{c|}{App} & \multicolumn{1}{c|}{Baseline} & \multicolumn{1}{c|}{App} & \multicolumn{1}{c|}{Baseline} & \multicolumn{1}{c|}{App} \\ \hline
P1                                       & \multicolumn{1}{l|}{0.5}      & \multicolumn{1}{l|}{0.8} & \multicolumn{1}{l|}{24.0}     & 30.0                     & \multicolumn{1}{l|}{1.0}      & \multicolumn{1}{l|}{1.0} & \multicolumn{1}{l|}{37.8}     & 31.8                     \\
P2                                       & \multicolumn{1}{l|}{1.0}      & \multicolumn{1}{l|}{0.8} & \multicolumn{1}{l|}{23.8}     & 23.3                     & \multicolumn{1}{l|}{1.0}      & \multicolumn{1}{l|}{1.0} & \multicolumn{1}{l|}{32.8}     & 32.5                     \\
P3                                       & \multicolumn{1}{l|}{0.3}      & \multicolumn{1}{l|}{0.8} & \multicolumn{1}{l|}{48.0}     & 33.7                     & \multicolumn{1}{l|}{1.0}      & \multicolumn{1}{l|}{1.0} & \multicolumn{1}{l|}{43.8}     & 35.8                     \\
P4                                       & \multicolumn{1}{l|}{1.0}      & \multicolumn{1}{l|}{1.0} & \multicolumn{1}{l|}{19.8}     & 18.5                     & \multicolumn{1}{l|}{1.0}      & \multicolumn{1}{l|}{1.0} & \multicolumn{1}{l|}{44.0}     & 33.5                     \\
P5                                       & \multicolumn{1}{l|}{1.0}      & \multicolumn{1}{l|}{1.0} & \multicolumn{1}{l|}{24.3}     & 23.5                     & \multicolumn{1}{l|}{1.0}      & \multicolumn{1}{l|}{1.0} & \multicolumn{1}{l|}{42.3}     & 37.8                     \\
P6                                       & \multicolumn{1}{l|}{0.8}      & \multicolumn{1}{l|}{1.0} & \multicolumn{1}{l|}{34.3}     & 38.0                     & \multicolumn{1}{l|}{0.8}      & \multicolumn{1}{l|}{1.0} & \multicolumn{1}{l|}{41.0}     & 40.0                     \\
P7                                       & \multicolumn{1}{l|}{1.0}      & \multicolumn{1}{l|}{1.0} & \multicolumn{1}{l|}{23.3}     & 21.3                     & \multicolumn{1}{l|}{1.0}      & \multicolumn{1}{l|}{1.0} & \multicolumn{1}{l|}{38.3}     & 33.5                     \\
P8                                       & \multicolumn{1}{l|}{0.0}      & \multicolumn{1}{l|}{0.5} & \multicolumn{1}{l|}{N/A}      & 28.0                     & \multicolumn{1}{l|}{0.5}      & \multicolumn{1}{l|}{0.8} & \multicolumn{1}{l|}{54.0}     & 39.7                     \\
P9                                       & \multicolumn{1}{l|}{0.0}      & \multicolumn{1}{l|}{0.5} & \multicolumn{1}{l|}{N/A}      & 29.5                     & \multicolumn{1}{l|}{0.8}      & \multicolumn{1}{l|}{1.0} & \multicolumn{1}{l|}{46.0}     & 33.8                     \\
P10                                      & \multicolumn{1}{l|}{1.0}      & \multicolumn{1}{l|}{1.0} & \multicolumn{1}{l|}{31.3}     & 23.0                     & \multicolumn{1}{l|}{1.0}      & \multicolumn{1}{l|}{1.0} & \multicolumn{1}{l|}{34.8}     & 32.8                     \\
P11                                      & \multicolumn{1}{l|}{0.5}      & \multicolumn{1}{l|}{1.0} & \multicolumn{1}{l|}{39.0}     & 39.5                     & \multicolumn{1}{l|}{0.5}      & \multicolumn{1}{l|}{0.8} & \multicolumn{1}{l|}{33.0}     & 40.3                     \\
P12                                      & \multicolumn{1}{l|}{0.5}      & \multicolumn{1}{l|}{1.0} & \multicolumn{1}{l|}{41.0}     & 35.5                     & \multicolumn{1}{l|}{0.5}      & \multicolumn{1}{l|}{0.8} & \multicolumn{1}{l|}{36.7}     & 30.3                     \\ \hline
\textbf{Mean}                            & \multicolumn{1}{l|}{0.6}      & \multicolumn{1}{l|}{0.9} & \multicolumn{1}{l|}{30.9}     & 28.6                     & \multicolumn{1}{l|}{0.8}      & \multicolumn{1}{l|}{0.9} & \multicolumn{1}{l|}{40.3}     & 35.1                     \\ \hline
\textbf{SD}                              & \multicolumn{1}{l|}{0.4}      & \multicolumn{1}{l|}{0.2} & \multicolumn{1}{l|}{9.4}      & 6.9                      & \multicolumn{1}{l|}{0.2}      & \multicolumn{1}{l|}{0.1} & \multicolumn{1}{l|}{6.1}      & 3.5                      \\ \hline
\end{tabular}
}
\caption{\label{tab:user-result} User study completion time and success rate for the two tasks. Each participant tries to complete the given task with and without our system for 4 times, respectively. Each user’s results shown here are averaged across the 4 trials. Completion times' averages are calculated only on successful trials. }
\end{table}